\documentclass[fleqn,usenatbib]{mnras}

\usepackage{newtxtext,newtxmath}

\usepackage[T1]{fontenc}

\usepackage{graphicx}	
\usepackage{amsmath}	
\usepackage{amssymb}	
\usepackage{multicol}   
\usepackage{pdflscape}	
\usepackage{graphicx}	
\usepackage{amsmath}	
\usepackage{amssymb}	
\usepackage{threeparttable,lscape}
\usepackage{natbib}
\usepackage{rotating}
\usepackage{color}
\usepackage{xcolor}
\usepackage{appendix}
\usepackage{multirow, array}
\usepackage{longtable}
\usepackage{lineno}
\usepackage{bbding}
\usepackage{pifont}
\usepackage{ae,aecompl}
\usepackage[normalem]{ulem}

\newcommand{\Msun}{M$_\odot$}
\newcommand{\Zsun}{Z$_\odot$}
\newcommand{\nodata}{\centering\arraybackslash --} 
\newcommand{\nifs}{\ensuremath{^{56}}Ni}
\newcommand{\cofs}{\ensuremath{^{56}}Co}
\newcommand{\kms}{km~s$^{-1}$}
\newcommand{\ergs}{erg~s^{-1}}

\definecolor{yaleblue}{rgb}{0.1,0.3,0.9}
\definecolor{ultramarine}{rgb}{0, 0, 150}
\definecolor{bostonuniversityred}{rgb}{0.8, 0.0, 0.0}
\definecolor{lava}{rgb}{0.81, 0.06, 0.13}
\definecolor{forestgreen}{rgb}{0.0, 0.27, 0.13}
\hypersetup{colorlinks=true, linkcolor=lava, urlcolor=yaleblue, citecolor=yaleblue}

\defcitealias{Jerkstrand15}{J15b}

\title[SN~2017ivv: IIb or not IIb?]{SN~2017ivv: two years of evolution of a transitional Type II supernova}

\author[C. P. Guti\'errez et al.]{
\parbox{\textwidth}{
\Large
C.~P.~Guti\'errez,$^{1}$\thanks{E-mail: C.P.Gutierrez-Avendano@soton.ac.uk}, 
A.~Pastorello,$^{2}$
A.~Jerkstrand,$^{3,4}$
L.~Galbany,$^{5}$
M.~Sullivan,$^{1}$
J.~P.~Anderson,$^{6}$
S.~Taubenberger,$^{7}$
H.~Kuncarayakti,$^{8,9}$
S.~Gonz\'alez-Gait\'an,$^{10}$
P.~Wiseman,$^{1}$
C.~Inserra,$^{11}$
M.~Fraser,$^{12}$
K.~Maguire,$^{13}$
S.~Smartt,$^{14}$ 
T.~E.~M\"uller-Bravo,$^{1}$
I.~Arcavi,$^{15,16}$ 
S.~Benetti$^{17}$
D.~Bersier,$^{18}$ 
S.~Bose,$^{19,20}$
K.~A.~Bostroem,$^{21}$
J.~Burke,$^{22,23}$ 
P.~Chen,$^{24,25}$ 
T.-W.~Chen,$^{7,26}$
M.~Della~Valle,$^{27,28}$ 
Subo~Dong,$^{24}$
A.~Gal-Yam,$^{29}$
M.~ Gromadzki,$^{30}$
D.~Hiramatsu,$^{22,23}$
T.~W.-S.~Holoien,$^{31}$\thanks{Carnegie Fellow}
G.~Hosseinzadeh,$^{32}$
D.~A.~Howell,$^{22,23}$
E.~Kankare,$^{8}$
C.~S.~Kochanek,$^{20,19}$
C.~McCully,$^{22,23}$
M.~Nicholl,$^{33,34}$
G.~Pignata,$^{35,36}$
J.~L.~Prieto,$^{37,36}$
B.~Shappee,$^{38}$
K.~Taggart,$^{18}$
L.~Tomasella$^{17}$
S.~Valenti,$^{21}$
D.~R.~Young$^{14}$ 
}
\vspace{0.4cm}
\\
\parbox{\textwidth}{Affiliations are listed at the end of the paper}
}

\date{Accepted XXX. Received YYY; in original form ZZZ}

\pubyear{2020}

\begin{document}
\label{firstpage}
\pagerange{\pageref{firstpage}--\pageref{lastpage}}
\maketitle

\begin{abstract}
We present the photometric and spectroscopic evolution of the type II supernova (SN~II) SN~2017ivv (also known as ASASSN-17qp). Located in an extremely faint galaxy (M$_r=-10.3$ mag), SN~2017ivv shows an  unprecedented evolution during the two years of observations. At early times, the light curve shows a fast rise ($\sim6-8$ days) to a peak of ${\rm M}^{\rm max}_{g}= -17.84$ mag, followed by a very rapid decline of $7.94\pm0.48$ mag per 100 days in the $V-$band. The extensive photometric coverage at late phases shows that the radioactive tail has two slopes, one steeper than that expected from the decay of $^{56}$Co (between 100 and 350 days), and another slower (after 450 days), probably produced by an additional energy source. From the bolometric light curve, we estimated that the amount of ejected $^{56}$Ni is $\sim 0.059\pm0.003$ M$\odot$. The nebular spectra of SN~2017ivv show a remarkable transformation that allows the evolution to be split into three phases: (1) H$\alpha$ strong phase ($<200$ days); (2) H$\alpha$ weak phase (between 200 and 350 days); and (3) H$\alpha$ broad phase ($>500$ days). We find that the nebular analysis favours a binary progenitor and an asymmetric explosion. Finally, comparing the nebular spectra of SN~2017ivv to models suggests a progenitor with a zero-age main-sequence mass of 15 -- 17 \Msun.
\end{abstract}



\section{Introduction}

Core collapse supernovae (CC-SNe) are produced by the explosion of massive stars ($>8$ \Msun). Observationally, they are a heterogeneous class, showing a large diversity in both spectra and photometry. They are broadly separated into objects with hydrogen in the spectra (type II SNe; SNe~II), and objects with no evidence of hydrogen lines \citep[SNe~I;][]{Minkowski41}. These hydrogen-poor SNe are classified as type Ib or type Ic SNe (SNe~Ib/c) based on the presence or absence of helium lines respectively \citep{Filippenko97, GalYam17}. 

Based on light curve shapes, SNe~II were initially subclassified into SNe with linear fast-declining light curves (SNe~IIL), and SNe with a slow decline or a plateau \citep[SNe~IIP;][]{Barbon79}. However, this separation has been refined with the analysis of larger samples of events that show a continuum in their photometric \citep[e.g.][]{Anderson14, Sanders15, Gonzalez15, Valenti16, Galbany16, Rubin16} and spectrocopic properties  \citep[e.g.][]{Gutierrez14,Gutierrez17b}, suggesting that all these events come from a similar progenitor population.

Transitional objects between SNe~II and SNe~Ib were identified with the discovery of SN~1987K \citep{Filippenko88}. This new class, identified as SNe~IIb \citep{Filippenko93}, encompasses objects showing broad hydrogen features in their spectra at early epochs consistent with a SN~II, but showing prominent helium lines at later phases, like SNe~Ib. This transitional class is thought to arise from stars that have lost most, but not all, of their hydrogen envelopes prior to explosion. How this happens is not fully understood, but it may occur via binary interaction \citep[e.g.,][]{Podsiadlowski92} or mass-loss through stellar winds \citep[e.g.,][]{Heger03,Puls08}.

The direct identification of nearby CC-SN progenitors in pre-explosion images has revealed that SNe~IIP arise from red supergiant (RSG) stars with masses of 8--18\Msun \citep[e.g.][]{VanDyk03, Smartt04, Smartt09, Smartt15}, while fast-declining objects (SN~IIL) may arise from more massive progenitors or binary systems (\citealt{Elias-Rosa10}, but see \citealt{Maund15}). The direct detection of the progenitor of SN~1993J \citep[e.g.][]{VanDyk02, Maund04, Maund09} and SN~2011dh \citep[e.g.][]{VanDyk11a, Maund11} suggests that many  SNe~IIb arise from stars in binary systems \citep[see also][]{Bersten12, Benvenuto13}. 

Within the single progenitor star scenario, a continuum between SNe~IIL and IIb is predicted \citep{Nomoto96,Heger03,Bayless15}; however, observational studies show some disagreements. \citet{Arcavi12} analysed the $R$-band light curve shape of 15 SNe~II and found a  subdivision into three classes: IIP, IIL, and IIb, which were interpreted as different progenitor populations. Later on, \citet{Faran14b}, analysing a sample of fast-declining SNe~II, found that SNe~IIL seem to be photometrically related to SNe~IIb. More recently, \citet{Pessi19}, using a larger sample of SNe (73 SNe~II and 22 SNe~IIb), found a lack of events that bridge the observed properties of these two classes, concluding that they form two observationally distinct families. 

Theoretical work \citep[e.g.,][]{Heger03} has shown that metallicity plays an important role in the evolution and final fate of massive stars. At low metallicity, the mass loss in massive stars is diminished  \citep[e.g.,][]{Woosley02} and the fraction of SNe~IIP is higher \citep[e.g.,][]{Heger03, Woosley15, Sravan19}. AT increased metallicity, mass loss reduces the hydrogen envelope and only SNe~IIL/IIb result \citep{Heger03}. However, for single RSG stars, some studies \citep[e.g.][]{vanLoon05,Goldman17,Chun18} suggest that the mass loss is independent of metallicity. If this is the case, the rate of RSG SN~IIb progenitors at low- and high-metallicity must be similar. However, recent results from SN host studies \citep{Galbany19} show that SNe~IIb environments are the most different from other CC-SNe: they occur in metal-poor, and relatively low star-formation rate environments (see also \citealt{Arcavi10}). \citet{Galbany19} argue that this result favours the binary system channel for SNe~IIb. 

While direct detections provide a relatively straightforward way to constrain the mass of the progenitor star, spectral modelling at late time ($>100-150$ days after explosion) can constrain the the nucleosynthesis yields and thus the zero-age main-sequence (ZAMS) mass of the progenitor \citep[][hereafter \citetalias{Jerkstrand15}]{Jerkstrand12, Jerkstrand14, Jerkstrand15}. During late phases, the SN ejecta becomes optically thin in the continuum and the stellar layers corresponding to the helium core are exposed. Stellar evolution models \citep[e.g][]{Woosley95} predict that the metal mass, particularly oxygen, strongly increases with progenitor ZAMS mass. 

Although spectral modeling provides insights into the properties of the progenitor, to date, only a few nebular phase statistical studies of SNe~II \citep{Maguire12,Jerkstrand15a,Valenti16,Silverman17} and SNe~IIb (\citealt{Ergon15}, \citetalias{Jerkstrand15}, \citealt{Fang18}) have been published. Individual analyses of SNe~IIb during the nebular phase \citep[e.g.][]{Matheson00, Taubenberger11} have shown a large diversity in this subclass, although a statistical characterization is still missing.

SNe~IIb are relatively rare events, representing 10$-$12 per cent of all CC-SNe \citep{Li11a,Shivvers17}, and only a few objects have been observed and analysed in detail. 
From these individual studies, SNe~IIb have shown a large variety in their photometric properties that allows us to divide them into two groups: SNe~IIb showing double-peaked light curves (e.g. SN~1993J, \citealt{Richmond94, Barbon95, Matheson00}; SN~2011fu,  \citealt{Kumar13,Morales-Garoffolo15}; SN~2016gkg \citealp{Arcavi17a, Bersten18}) and SNe~IIb with a single peak (e.g. SN~2008ax, \citealt{Pastorello08, Taubenberger11}; SN~2011dh \citealt{Arcavi11,Sahu13,Ergon14,Ergon15};  SN~2015as \citealt{Gangopadhyay18}). For the double-peak light curves, the initial peak, lasting less than two days, has been attributed to the energy radiated by the shock \citep[e.g.][]{Woosley94, Bersten12, Nakar14}, while the second peak, with a duration of about two weeks, is powered by the decay of $^{56}$Ni. The double-peaked group is thought to arise from extended massive stars that interact with the dense wind from the progenitor, while the single-peaked group arises from compact stars \citep{Chevalier10}.

In this paper, we present the analysis and results from the photometric and spectroscopic monitoring of SN~2017ivv over two years. SN~2017ivv exploded in the faintest host galaxy reported for a SN~II to date. 
Due to its exceptional behaviour, SN~2017ivv provides a great opportunity to explore and understand the connection and diversity within the SNe II and ~IIb classes, and the possible effects of metallicity in the SN evolution. 	

The paper is organized as follows. A description of the observations and data reduction are presented in Section~\ref{observations}. The host galaxy is characterised in Section~\ref{galaxy}. We describe the photometric and spectroscopic properties of SN~2017ivv in Section~\ref{photo} and  Section~\ref{spec}, respectively. Finally, in Section~\ref{analysis} we present the analysis and discussion. We provide our conclusions in Section~\ref{conclusions}.
{\noindent}Throughout, we assume a flat $\Lambda$CDM universe, with a Hubble constant of $H_0=70$\,km\,s$^{-1}$\,Mpc$^{-1}$, and
$\Omega_\mathrm{m}=0.3$.

\section{Observations of SN~2017ivv}
\label{observations}

\subsection{Discovery, explosion epoch and classification}

SN~2017ivv (also known as ASASSN-17qp and ATLAS17ntk) was discovered by the All-Sky Automated Survey for Supernovae \footnote{\url{http://www.astronomy.ohio-state.edu/asassn/index.shtml}}, ASAS-SN \citep{Shappee14,Kochanek17} on 2017 December 12 (MJD=58099.03; \citealt{Brimacombe17}) with a magnitude of $g\sim14.5$ mag. It was spectroscopically classified as a SN~II by the Asiago Transient Classification Program \citep{Tomasella14} on 2017 December 15 (MJD=58102.73). The last non-detection was obtained on 2017 December 4 (MJD=58091.03), with a detection limit of $g\sim17.14$ mag. An early detection in the orange filter ($o=15.26$ mag) was obtained on 2017 December 5 (MJD=58092.22) by the Asteroid Terrestrial-impact Last Alert System (ATLAS; \citealt{Tonry18, Smith20}). Given the first detection from ATLAS and last non-detection from ASASSN, we adopt 2017 December 4 (MJD=$58091.62\pm0.59$) as the explosion date.  

SN~2017ivv was classified as a SN~II based on the presence of hydrogen lines in the spectra. As part of the extended Public ESO Spectroscopic Survey of Transient Objects (ePESSTO; \citet{Smartt15a}), we started a follow-up campaign through the program \lq SNe~II in low-luminosity host galaxies\rq. To check the initial classification, the Supernova Identification (\textsc{snid}, \citealt{Blondin07}) code and the GEneric cLAssification TOol (\textsc{Gelato}; \citealt{Harutyunyan08}) were used and good matches for SN~2017ivv were found with several SNe~II (e.g. SN~2007od, \citealt{Inserra11}; SN~1995ad, \citealt{Inserra13}) and a few SNe~IIb (e.g. SN~2011fu, \citealt{Kumar13, Morales-Garoffolo15}). A H$\alpha$ P-Cygni profile with a broad emission component, but a weak absorption component suggested it is a fast-declining SN~II. However, a gap in the observations between 20 and 100 days does not allow the study of the appearance of spectral lines and light curve evolution during the recombination phase. When the SN became visible again, at 111 days from explosion, the spectra and light curve are still similar to fast-declining SNe~II. An analysis of the emission lines in the nebular spectra after 200 days suggests SN~2017ivv is not a typical hydrogen-rich SN. It seems to have a lower hydrogen mass than any other object of this type. As SNe~IIb are assumed to have much less hydrogen, we compare SN~2017ivv with both fast-declining SNe~II and SNe~IIb.

\subsection{Optical photometry}

Optical imaging of SN~2017ivv was acquired with several telescopes and instruments (Tables~\ref{photoatlas} -- \ref{photomuse}). Details of the observation are briefly summarised here.

\begin{itemize}

\item \textbf{ATLAS:} Photometry in the \textit{orange (o)} filter (a red filter that covers \textit{r+i}) was obtained by the twin 0.5-m ATLAS telescope system \citep{Tonry18}, from 2017 December 5 to 2018 October 15. The ATLAS data were reduced and calibrated automatically as described in \citet{Tonry18} and \citet{Smith20}. The mean magnitudes are reported in  Table~\ref{photoatlas}. 

\item \textbf{Las Cumbres Observatory:} Most of the multiband optical photometry was obtained with the 1.0-m telescopes of Las Cumbres Observatory \citep{Brown13}, through the ePESSTO and the Global Supernova Project allocated time. The photometric monitoring began on 2018 March 16, three months after the discovery and as soon as the SN was visible after being sun-constrained. An early observation in the $BVri$-bands was obtained on 2017 December 14. All data reduction was performed following the prescriptions described by \citet{Firth15}. The $BgVri$ magnitudes of SN~2017ivv are presented in Table~\ref{photolco}.
 
\item \textbf{Asiago:} Six epochs of $uBgVriz$ photometry were obtained with the Copernico 1.82 m telescope equipped with AFOSF and the 67/91 Schmidt Telescope at the Asiago Observatory. The images were reduced following standard procedures (including bias, dark, and flat-field corrections). The $ugriz$ magnitudes were calibrated using observations of local Sloan and Pan-STARRS sequences. The $BV$ magnitudes were derived using Pan-STARRS and the transformations in \citet{Chonis08}. Table~\ref{photoasiago} presents the Asiago photometry.

\item \textbf{ASAS-SN:} $V$ and $g$-band photometric observations were obtained with the ASAS-SN units `Brutus' (Hawaii), `Cassius' (CTIO, Chile) and `Henrietta Leavitt' (McDonald, Texas).  All images were processed with an automated pipeline \citep{Shappee14} using the ISIS image subtraction package \citep{Alard98, Alard00}. Details of the process are presented in \citet{Shappee14}. We performed aperture photometry on the subtracted images using the IRAF \textit{apphot} package, and then calibrated the results using the AAVSO Photometric All-Sky Survey \citep[APASS;][]{Henden15}. The ASAS-SN photometry is presented in Table~\ref{photoasas}. 

\item \textbf{Post Observatory SRO:} Five epochs of $BVri$ photometry were acquired with the Apogee Alta U230 camera and the Apogee Alta U47 with 0.6 m telescopes at Post Observatory SRO (CA, USA) and Mayhill (NM, USA), respectively. All images were processed following standard procedures. Table~\ref{photopost} presents the Post photometry.

\item \textbf{GROND:} Two epochs of $r-$band and five epochs of $JHK$ photometry were obtained with the Gamma-Ray Burst Optical/Near-Infrared Detector (GROND; \citealt{Greiner08}) on the 2.2-m MPG telescope at the European Southern Observatory (ESO) La Silla Observatory in Chile. The images were reduced with the GROND pipeline \citep{Kruhler08}. The $r$ magnitudes were calibrated using Pan-STARRS, while the $JHK$ magnitudes were calibrated using 2MASS. The photometry is presented in Table~\ref{photogrond}.

\item \textbf{MUSE:}  Two epochs of synthetic $Vri-$band photometry were extracted from the spectrum of SN~2017ivv obtained by the Multi Unit Spectroscopic Explorer (MUSE; \citealt{Bacon14}) at the ESO Very Large Telescope (VLT). The MUSE data were reduced using the \texttt{EsoReflex} pipeline \citep{Freudling13}. 
Table~\ref{photomuse} presents the MUSE synthetic photometry.

\end{itemize}

\subsection{Optical spectra}

Spectroscopic observations of SN~2017ivv were acquired with six different instruments: the ESO Faint Object Spectrograph and Camera (EFOSC2; \citealt{Buzzoni84}) at the 3.5-m ESO New Technology Telescope (NTT), the Asiago Faint Object Spectrograph and Camera (AFOSC) at the 1.82-m Copernico Telescope, the  Boller and Chivens spectrograph (BC) at the 1.22-m Galileo Telescope, the FOcal Reducer/low dispersion Spectrograph 2 (FORS2; \citealt{Appenzeller98}), MUSE at the VLT, and the Low Resolution Imaging Spectrometer (LRIS; \citealt{Oke95}) at the 10-m Keck telescope. In total, we obtained 25 epochs spanning between 10 and 671 days from explosion. Details of the instruments used for the spectral observations are reported in Table~\ref{tspectra}.

Data reductions for AFOSC and BC were performed with \textsc{iraf} using standard routines (bias subtraction, flat-field correction, 1D extraction, and wavelength and flux calibration). For EFOSC2, the data were reduced using the PESSTO pipeline \citep{Smartt15a}. The spectrum obtained with FORS2 and the cube observed with MUSE were reduced using the \textsc{EsoReflex} pipeline \citep{Freudling13}. 
The Keck LRIS spectrum was reduced with the \textsc{lpipe} \citep{Perley19} pipeline using default parameters and the standard reduction procedures. Finally, we calibrated the absolute flux of the spectra by using the $r-$band magnitudes to match the photometry.
All spectra are available through the WISeREP\footnote{\url{http://wiserep.weizmann.ac.il/home}} archive \citep{Yaron12}.

\section{The host galaxy}
\label{galaxy}

\begin{figure}
\centering
\includegraphics[width=0.85\columnwidth]{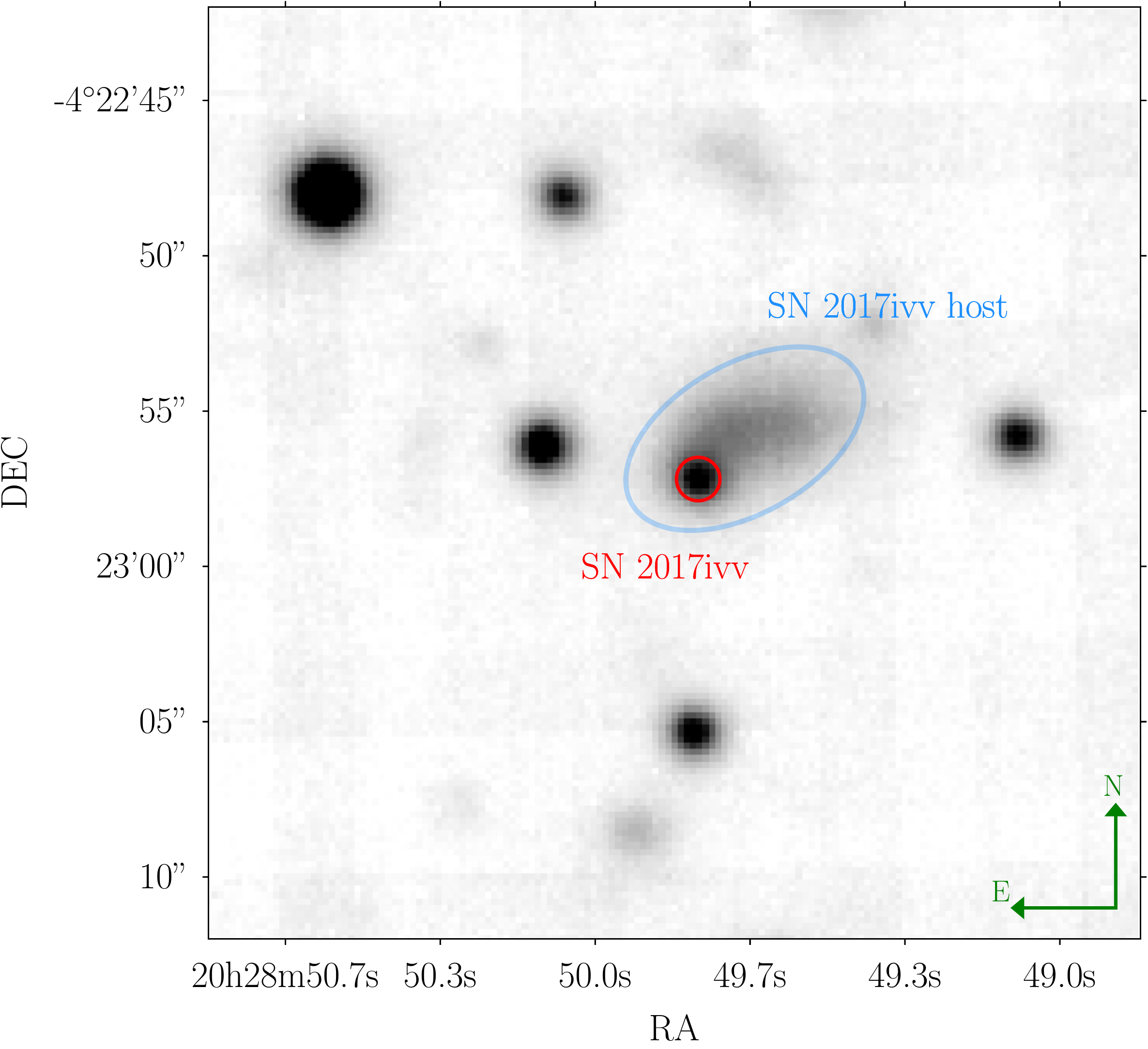}
\includegraphics[width=0.95\columnwidth]{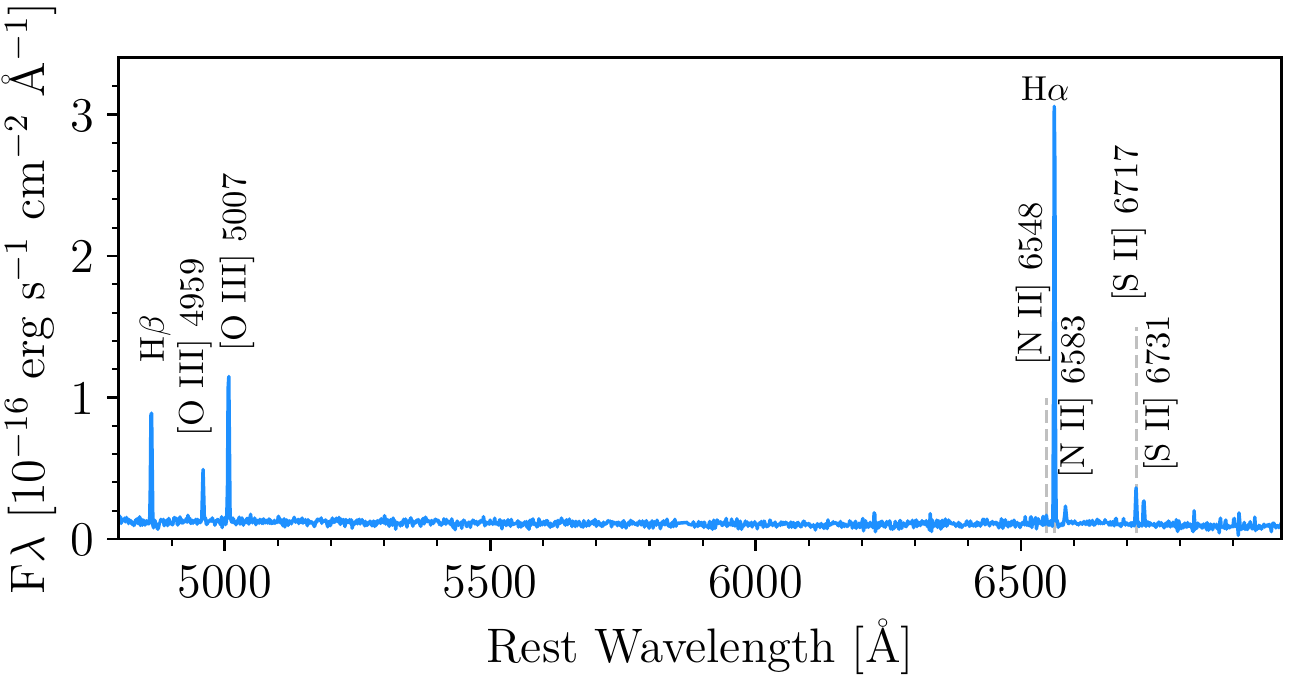}
\caption{\textbf{Top panel:} MUSE image of SN~2017ivv and its host galaxy. This synthetic image is created by collapsing the data cube across the full wavelength range of the observation. Orientation is North-up, East-left. The red circle marks the SN, while the sky-blue ellipse marks its host galaxy. \textbf{Bottom panel:} MUSE optical spectrum for the \ion{H}{II} region of SN~2017ivv host. The brightest optical emission lines are labeled. Observations were obtained on 2019 May 11.}
\label{gal}
\end{figure}

The host galaxy of SN~2017ivv is GALEXASC J202849.46-042255.5. It is a small and very faint galaxy with no previously published redshift. The only available information comes from a detection in Pan-STARRS 3Pi at a Kron mag of $g=21.27\pm0.06$ mag \citep{Flewelling16}. 
We obtained MUSE observations on 2019 May 11 and October 10 as a part of the All-weather MUse Supernova Integral field Nearby Galaxies (AMUSING, \citealt{Galbany16a}). The observations were taken at around 520 and 671 days from explosion and the SN was still detected. Figure~\ref{gal} (top panel) shows the synthetic $r-$band image of the host galaxy of SN~2017ivv at 520 days. In the bottom panel, the spectrum of the \ion{H}{II} region near SN~2017ivv position is shown. The spectrum reveals emission lines of H$\alpha$, H$\beta$, \ion{[O}{III]} $\lambda4959$, \ion{[O}{III]} $\lambda5007$, \ion{[N}{II]} $\lambda6548$, \ion{[N}{II]} $\lambda6583$, \ion{[S}{II]} $\lambda6717$, and \ion{[S}{II]} $\lambda6731$ at a redshift of 0.0056. Due to the lack of independent measurements of distance, we assume that the recession velocity is all due to cosmological expansion. However, to estimate the uncertainty in the measurements, we use a peculiar velocity of $\pm200$ km s$^{-1}$ \citep{Tully13}. With these values, we compute a distance of $d=24.09\pm2.90$ Mpc, which corresponds to a distance modulus of $\mu=31.91^{+0.25}_{-0.28}$ mag.

The Galactic reddening in the direction of SN~2017ivv is $E(B-V)=0.05$ mag \citep{Schlafly11}. Due to the absence of narrow interstellar \ion{Na}{I} D lines ($\lambda\lambda5889$, 5895) in the SN spectra (see Section~\ref{spec}) and the low luminosity of the host, we assume negligible host reddening.   
Given the synthetic $r-$band magnitude measured from the MUSE spectrum ($r=21.74\pm0.19$ mag), the Galactic reddening and the distance modulus, we obtain a host galaxy absolute $r-$band magnitude of $-10.3$ mag (M$_g=-10.8$ mag, using the Pan-STARRS magnitude). This extremely low luminosity makes GALEXASC J202849.46$-$042255.5 the faintest host for a SN~II to date. 
Using the spectrum of the \ion{H}{II} region near SN~2017ivv, we estimate the oxygen abundance and the star-formation rate (SFR). Measuring the fluxes of H$\alpha$, H$\beta$, \ion{[O}{III]} $\lambda5007$ and \ion{[N}{II]} $\lambda6583$ and  applying the O3N2 and N2 diagnostic methods from \citet{Marino13}, we obtain an oxygen abundance of 12 + log(O/H)$=8.247\pm0.066$ dex and 12 + log(O/H)$=8.131\pm0.056$ dex, respectively. 
With these values, the host galaxy of SN~2017ivv is in the lowest $\sim$3 and $\sim$17 per cent of the sample analysed by \citet{Anderson16}. 

The SFR can be calculated using the extinction corrected luminosity of the H$\alpha$ emission, L(H$\alpha$), and  the normalisations from  \citet{Kennicutt12}. For  L(H$\alpha$)$=1.089\times10^{38}$ erg s$^{-1}$, we derive $\mathrm{SFR}=5.85\times10^{-4}$ \Msun yr$^{-1}$ ($\log\left(\mathrm{SFR}\right) = -3.05$ \Msun yr$^{-1}$), which is much smaller than the average value of the Milky Way ($SFR=1.9\pm0.4$ \Msun yr$^{-1}$, \citealt{Chomiuk11}).

We obtained $grizy$ photometry of the host galaxy from the Pan-STARRS1 (PS1) public science archive\footnote{\url{catalogs.mast.stsci.edu/panstarrs/}}. To obtain estimates of the stellar mass ($M_*$) and SFR, we fit the broadband SED from this photometry with templates from the single stellar population models of \citet{Bruzual03} via the \textsc{CIGALE} code \citep{Boquien19}. We assume a \citet{Chabrier03} initial mass function (IMF). The best fitting templates correspond to $M_* = \log \left(M/M_{\odot}\right) = 5.93\pm 0.27$ and $\log\left(\mathrm{SFR}\right) = -3.22 \pm 0.2$ \Msun yr$^{-1}$. While these uncertainties are likely underestimated, we note that the SFR matches that measured from the H$\alpha$ luminosity. This SFR corresponds to a specific SFR (sSFR) of $\log\left(\mathrm{sSFR/yr}^{-1}\right) = -9.16 ^{0.46}_{0.48}$, which is typical of star-forming galaxies and other SN II hosts  (i.e. this is not a star-burst; see, e.g., \citealt{Taggart19}).

\section{Photometric properties}
\label{photo}

\subsection{Light curve evolution}
\label{slc}

\begin{figure*}
\centering
\includegraphics[width=11.2cm]{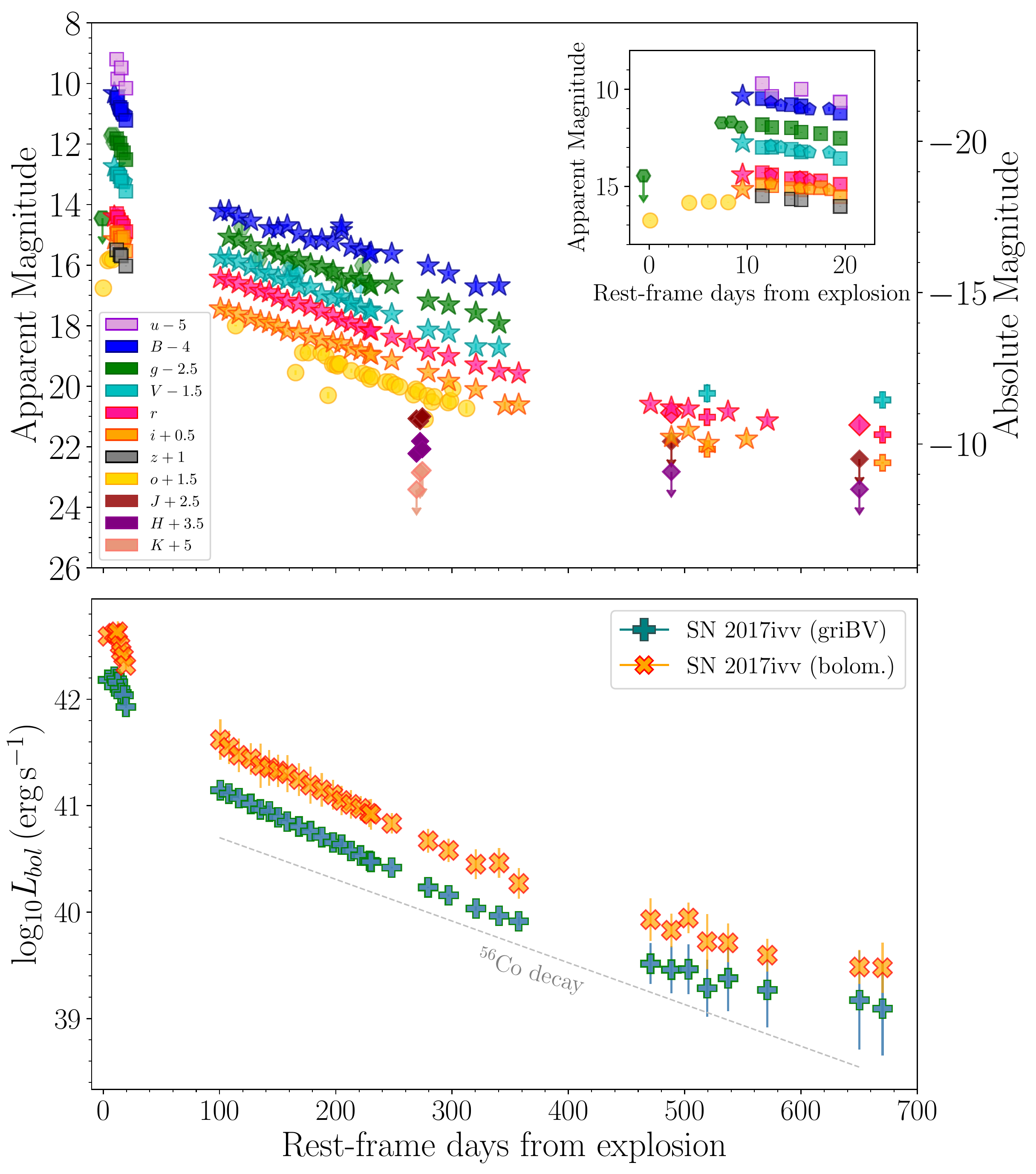}
\caption{\textbf{Upper:} Optical light curves of SN~2017ivv. ATLAS photometry is presented as circles, Las Cumbres Observatory photometry as stars, Asiago photometry as squares, ASAS-SN photometry as hexagons, Post photometry as pentagons, GROND photometry as diamonds, and MUSE synthetic photometry as crosses. The photometry is corrected for MW extinction. Note that the transition between the photospheric and nebular phase happened when the supernova was behind the Sun. The inset shows the light curves around the peak. \textbf{Bottom:} Bolometric (orange) and pseudo-bolometric (dark cyan) light curves of SN~2017ivv.}
\label{lc}
\end{figure*}

The multiband light curves of SN~2017ivv are presented in Figure~\ref{lc}. A high cadence over the first 20 days allows us to constrain the magnitude at maximum and the rise time. Using a polynomial fit, we find that the maximum occurs at $\sim6$ ($o$-band) and 8 days ($g-$band) from the explosion at ${\rm M}^{\rm max}_{o}= -17.63$ mag and ${\rm M}^{\rm max}_{g}= -17.84$ mag, respectively. This rise is consistent with the distribution of fast rise-times ($7.5\pm0.3$ days) found by \citet{Gonzalez15} for hydrogen-rich SNe. After the peak, a rapid decline is observed in all bands. Following the prescriptions of \citet{Anderson14}, we find a decline rate of $7.94\pm0.48$ mag (100 d)$^{-1}$ in the $V-$band. Comparing this decline rate with values from the literature, we find that SN~2017ivv has the fastest initial decline measured of any SN~II \citep{Gutierrez17a}, but it is a typical decline rate for a SN~IIb  (5 to 9 mag (100 d)$^{-1}$ after peak).

Unfortunately, due to a gap in observations between $20-100$ days, the transition from the photospheric to the nebular phase was missed. The SN was observable again at around 100 days post explosion and observed in the $BgVri$ filters. The slope in the radioactive tail can be split into two phases, one between 100 and 350 days, and a second one after 400 days. For the first time period, we obtain a value of $1.36\pm0.02$ mag (100 d)$^{-1}$ in $V$ band and $1.33\pm0.01$ mag (100 d)$^{-1}$ in $r$ band. For the second interval, the decay is $0.14\pm0.13$ mag (100 d)$^{-1}$ in the $V$ band, and $0.34\pm0.10$ mag (100 d)$^{-1}$ in the $r$ band. The slope between 100 and 350 days is larger than expected for the full-trapping of gamma-ray photons from the decay of $^{56}$Co (0.98 mag per 100 days in the $V$ band; \citealt{Woosley89}), suggesting that SN~2017ivv has low ejecta mass \citep{Anderson14}, but after 400 days the slope is smaller, which could be caused by some interaction between the ejecta and  circumstellar material (CSM). 

The high decline rate in the photospheric phase suggests that SN~2017ivv could be either a fast-declining SN~II or a SNe~IIb. For this reason we will compare SN~2017ivv to both fast-declining SNe~II and SNe~IIb with well-sampled light curves and some spectra during both the photospheric and nebular phase. The fast-declining SN~II group consists of ASASSN-15oz \citep{Bostroem19}, SN~2014G \citep{Terreran16}, SN~2013ej \citep{Valenti14,Huang15,Yuan16}, and SN~2013by \citep{Valenti15,Black17}, while the SNe~IIb sample consists of SN~2015as \citep{Gangopadhyay18}, SN~2013df \citep{Morales-Garoffolo14,Szalai16}, SN~2011hs \citep{Bufano14}, SN~2011dh \citep{Arcavi11,Sahu13,Ergon14,Ergon15}, SN~2008ax \citep{Pastorello08, Taubenberger11,Modjaz14}, and SN~1993J \citep{Barbon95, Matheson00}. Because of the extremely low luminosity of the host of SN~2017ivv, we also include SN~2015bs \citep{Anderson18}, which is a slow-declining SN~II that exploded in the faintest host galaxy found before SN~2017ivv.

\begin{figure}
\centering
\includegraphics[width=0.99\columnwidth]{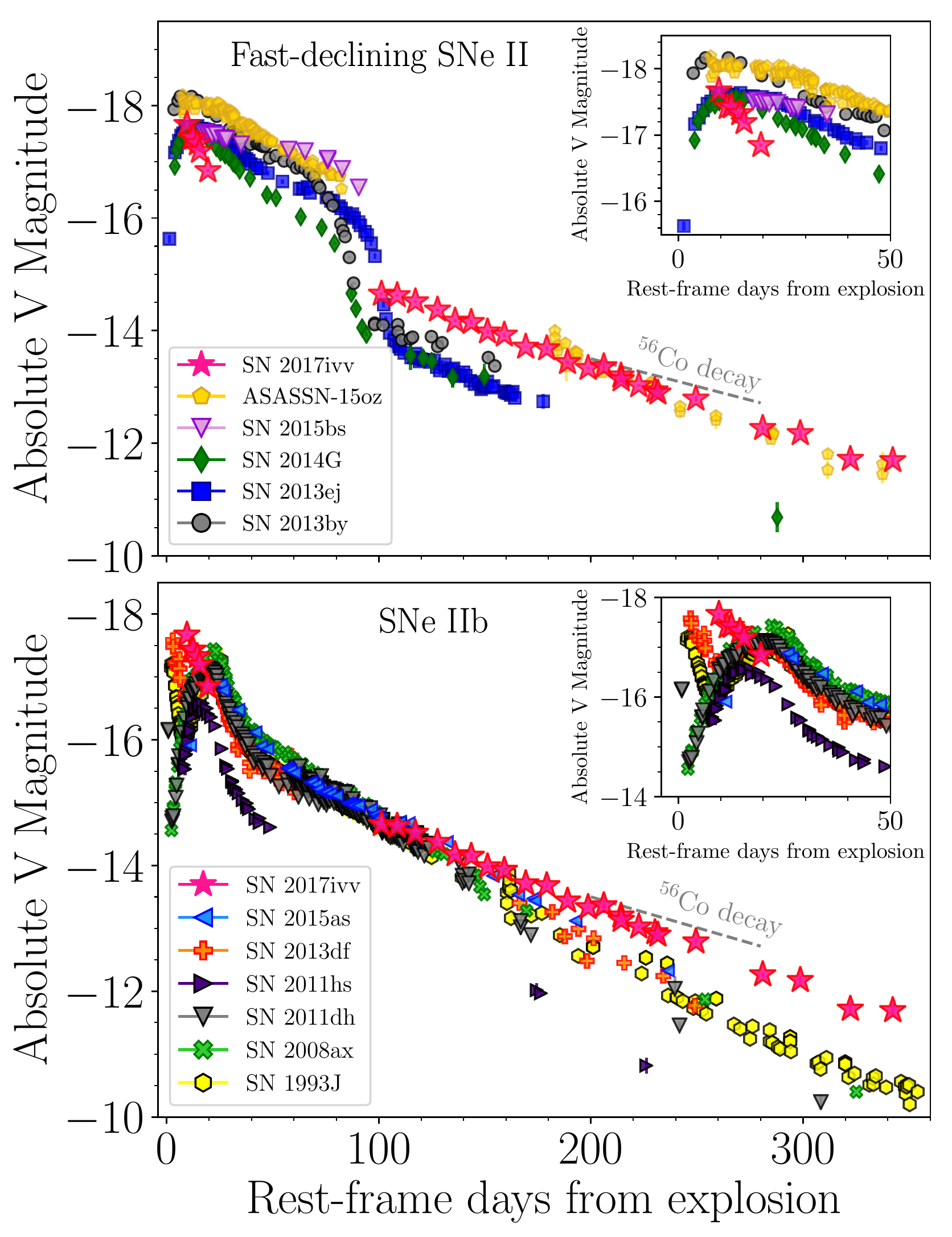}
\caption{Photometric comparison of SN~2017ivv with the fast-declining SN~II (top) and SN~IIb (bottom) samples. The extinction and distance assumptions to calculate the absolute magnitudes were adopted from the references in Section~\ref{slc}.}
\label{lc_comp}
\end{figure}

In Figure~\ref{lc_comp} (top panel), we compare the $V-$band light curve of SN~2017ivv with the type II sample. The comparison shows that SN~2017ivv has a peak luminosity similar to SN~2014G and SN~2013ej. Despite the gap during the photospheric phase, it is possible to measure the drop in luminosity from the peak to the nebular phase (100 days). For SN~2017ivv, the luminosity drops by $\lesssim3$ mag, while SN~2014G, SN~2013ej and SN~2013by decrease by more than 4 mag. During the nebular phase, SN~2017ivv has a luminosity comparable to ASASSN-15oz. In terms of slope, it has a similar decline rate to SN~2013ej and ASASSN-15oz. 

In the bottom panel of Figure~\ref{lc_comp}, the $V-$band light curve of SN~2017ivv is compared with SNe~IIb. Because of the good constraint on the explosion epoch (see Figure~\ref{lc}), we could confirm that the early narrow peak observed in SN~1993J and SN~2013df is not seen in SN~2017ivv (SN~2008ax does not have it either). This  peak has been attributed to the cooling of the ejecta after the shock breakout \citep[e.g.][]{Woosley94, Bersten12, Nakar14}. 
SN~2017ivv displays the shortest rise-time to peak ($\sim6-8$ days) and is also the most luminous object (M$_V=-17.67$ mag) among the SN~IIb comparison sample. Its luminosity is a slightly higher than SN~2008ax and SN~1993J, which peaked at M$_V=-17.61$ mag, and M$_V=-17.59$ mag, respectively. After 150 days, the decline rate of SN~2017ivv is slower than all SNe~IIb in the comparison sample. This slow decline indicates that the gamma escape in SN~2017ivv is less strong, implying a higher ejecta mass.

\subsection{Bolometric luminosity}
\label{bol}

To calculate the pseudo-bolometric luminosity, we use the extinction corrected $BgVri$ bands converted to fluxes at the effective wavelength for each filter. We integrate a spectral energy distribution (SED) over the wavelength range of the filters, assuming zero flux at the limits. The emitted fluxes were computed at epochs when $BgVri$ were observed simultaneously. When observations in a band were unavailable for a  given epoch, the magnitudes were obtained by interpolating or extrapolating the light curves using low-order polynomials or assuming constant colours. Fluxes were converted to luminosity using the distance adopted in Section~\ref{galaxy}. 
As an estimate of the full bolometric light curve, we extrapolated the SED constructed from the $BgVri$ bands using a blackbody fit to the SED. With a polynomial fit, we find a peak luminosity of L$_{bol}=4.4\times10^{42}$ erg s$^{-1}$ occurring at $\sim10.5\pm2$ days after explosion (MJD=58102.18). This maximum is about $\sim2.5-4.5$ days later than the peaks in the $o$ and $g-$bands. After peak, the light curve declines at a rate of 7.97 mag per 100 days, quite similar to the $V-$band. The slope in the tail is 1.31 mag (100 d)$^{-1}$, a slightly smaller than the $V-$band slope of  (1.36 mag (100 d)$^{-1}$). These rapid declines suggest that the gamma-ray trapping is incomplete.

The \nifs\ mass can estimated by comparing the bolometric luminosity of SN~2017ivv with the theoretical values of full \cofs\ deposition, as reported by \citet{Jerkstrand12}:

\begin{equation}
  L_0(t)=9.92\times10^{41} \frac{M(\nifs)}{0.07 M_{\odot}} (e^{-t/111.4}-e^{-t/8.8})~{\rm \ergs},
\end{equation}
 
{\noindent}where M(\nifs) is the mass of \nifs\ synthesised during the explosion. As the light curve tail does not follow the \cofs\ decay, the gamma escapes needs to be considered \citep{Clocchiatti97,Jerkstrand17} as follows:

\begin{equation}
  L(t)=L_0(t)\times(1-e^{-(\tau_{tr}/t)^2})~{\rm \ergs},
\end{equation}

{\noindent}where $L_0$ comes from equation (1) and $\tau_{tr}$ is the full-trapping characteristic time-scale. We note that equation (2) is appropriate as long as positron contributions are minor (likely before 350 d).
Using these equations to fit the bolometric light curve of SN~2017ivv between 120 and 350 days, we find a nickel mass of M$(\nifs)=0.059\pm0.003$ \Msun\ and $\tau_{tr}=298\pm8$ days.

\subsection{Evolution of the colour curves}
\label{scol}

\begin{figure}
\centering
\includegraphics[width=\columnwidth]{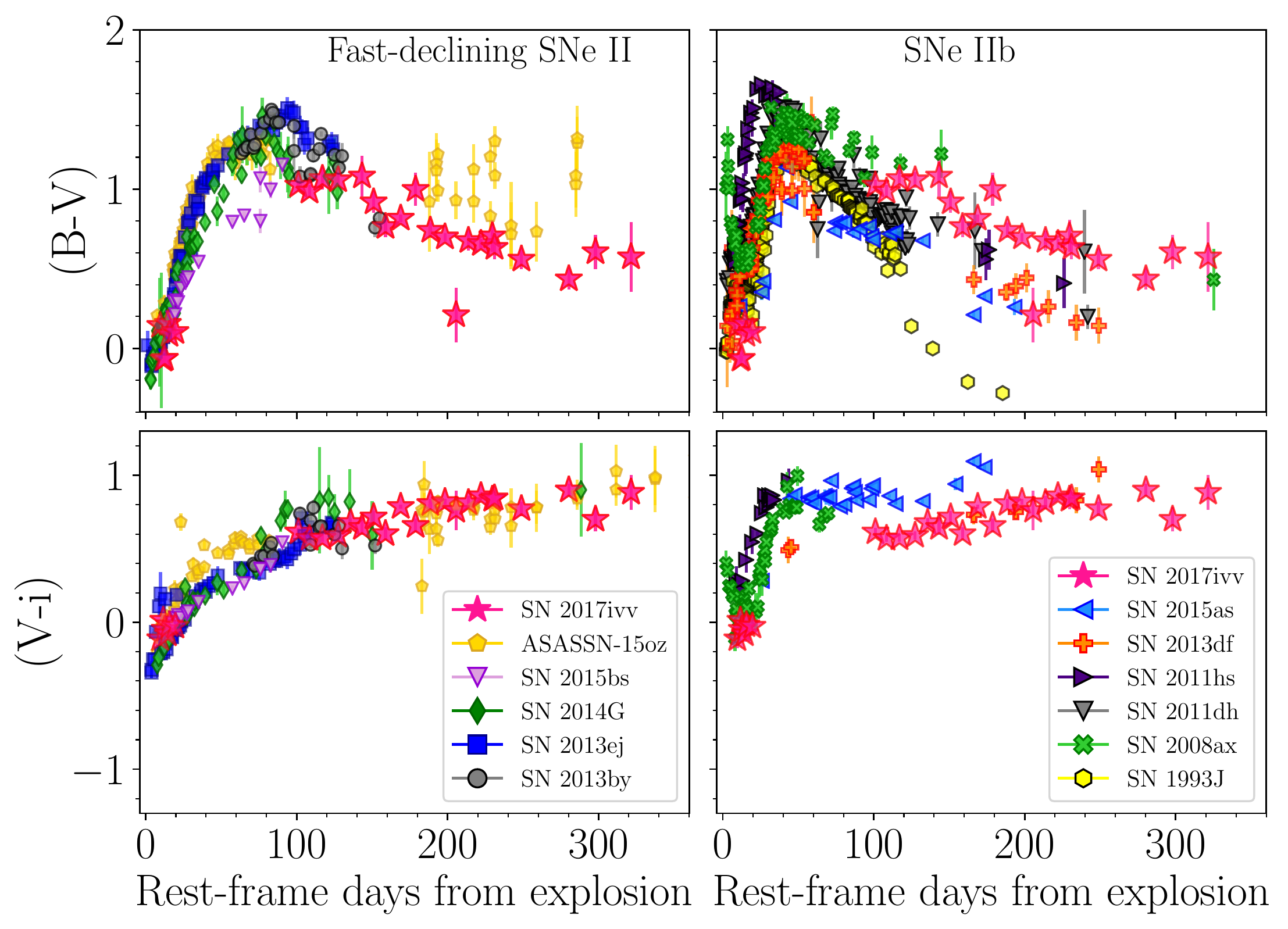}
\caption{Colour evolution of SN~2017ivv compared to fast-declining SNe~II (left panels) and SNe~IIb (right panels). See Section~\ref{slc} for the references.}
\label{col}
\end{figure}

Figure~\ref{col} compares the $(B-V)$ and $(V-i)$ colours of SN~2017ivv with the fast-declining sample and SN~2015bs (left panel) and with SNe~IIb (right panel). In the former, the behaviour of all SNe at early phases is quite similar. At late phases, the colour evolution of SN~2017ivv is comparable to ASASSN-15oz. Unfortunately, the other objects have no colour information after 100 days. Nevertheless, the overall colour evolution of SN~2017ivv seems similar to that observed in the fast-declining objects.

SN~2017ivv is bluer at early times than the SNe~IIb sample. Then, it rapidly becomes redder. Between 100 and 150 days, the $(B-V)$ evolution is nearly similar to that of SN~2008ax, although it is redder than most of the comparison sample. After 150 days, it evolves back to bluer colours. At epochs later than 100 days, the $(V-i)$ colours of SN~2017ivv are similar to those of SN~2013df.

\section{Spectral properties}
\label{spec}

\subsection{Early phases}
\label{Searly}

\begin{figure}
\centering
\includegraphics[width=\columnwidth]{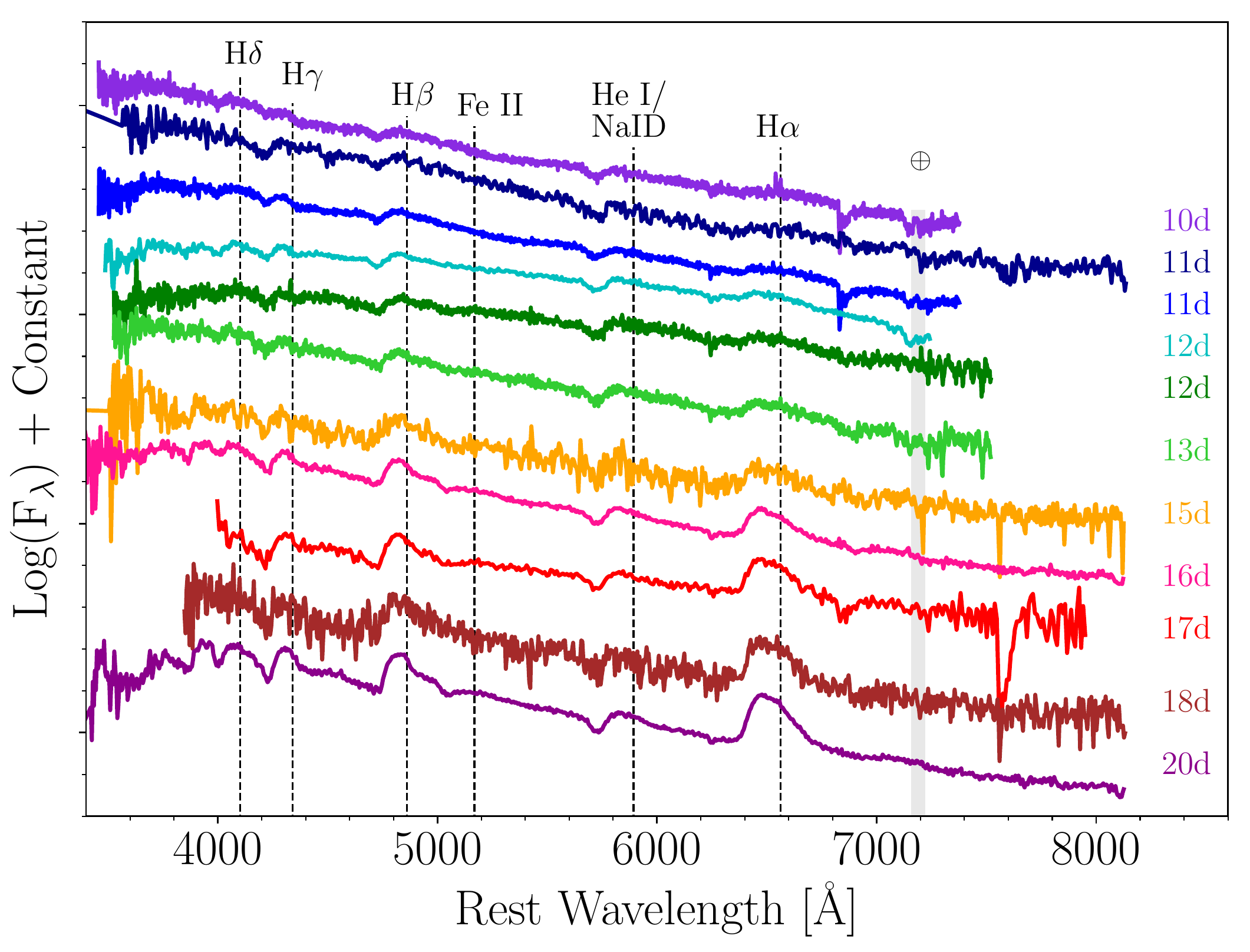}
\caption{Optical spectra of SN~2017ivv ranging from 10 to 20 days from explosion.
The phases are labeled on the right. Each spectrum has been corrected for
Milky Way reddening and shifted in flux for  comparison.}
\label{spectra}
\end{figure}

Figure~\ref{spectra} shows the optical spectra of SN~2017ivv from 10 to 20 days. The early spectra display a blue continuum, with low-contrast hydrogen and \ion{He}{I} $\lambda5876$ lines that are typical of young SNe~II. At 16 days, these lines become stronger, and broad P-Cygni profiles are visible. In particular, H$\alpha$ exhibits a  broad and asymmetric emission component with an extremely weak absorption feature. Weak absorption profiles have been seen in fast-declining SNe~II and interpreted as evidence for a low amount of absorbing material along the line of sight due to either a low-mass envelope or a very steep density gradient \citep[e.g.][]{Schlegel96, Gutierrez14}. Asymmetric emission profiles are also detected in H$\beta$. Between 16 and 20 days, the emission component of H$\alpha$ shows a flat-topped profile indicating either a weak interaction with a low-density CSM (e.g. \citealt{Inserra11}) or the presence of an expanding, geometrically thin H shell \citep{Barbon95}. Similar profiles have been seen before in well-studied SNe~II at various phases, e.g. SN~2007od, \citep{Inserra11, Gutierrez17a}; SN~2007X, \citep{Gutierrez17a}; SN~2016esw, \citep{deJaeger18}) and SNe~IIb, e.g. SN~1993J, \citep{Barbon95,Matheson00}; SN~2011hs, \citep{Bufano14}.
After 20 days, \ion{Fe}{II} $\lambda5169$ starts to be visible. No narrow \ion{Na}{I} D ($\lambda\lambda5889$, 5895) absorption features from either the host galaxy or the Milky Way are detected, suggesting little reddening in the direction of the SN.

Overall, the line profiles of SN~2017ivv display blue-shifted emission line peaks. In the case of H$\alpha$, the velocity offset of the emission component is $\sim3300$ km s$^{-1}$. For H$\beta$, the velocity offset is $\sim2000$ km s$^{-1}$, while for \ion{He}{I}/\ion{Na}{I}\,D, it is between 2200 and 2700 km s$^{-1}$. According to \citet{Anderson14a}, such blue-shifted velocity offsets are common properties of SN~II spectra and are caused by the density distribution of the ejecta.

\begin{figure}
\centering
\includegraphics[width=0.9\columnwidth]{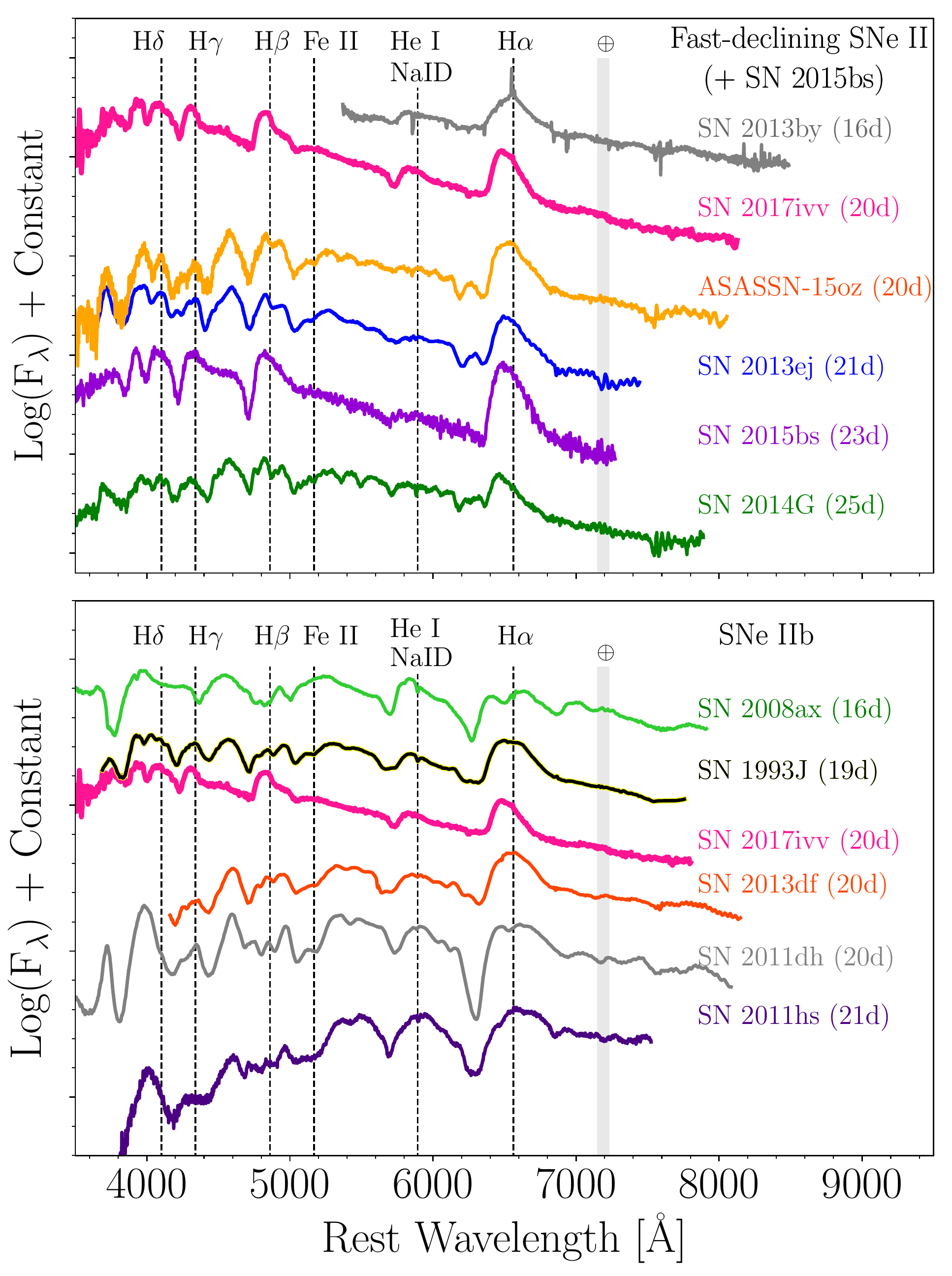}
\caption{Spectral comparison at $\sim20$ days of SN~2017ivv with the fast-declining SNe~II (top panel) and SNe~IIb (bottom panel). The spectra have been corrected by Milky Way reddening and redshift.}
\label{spcomp}
\end{figure}

In Figure~\ref{spcomp}, the spectrum of SN~2017ivv at 20 days from explosion is compared with the fast-declining SNe~II (top panel) and SNe~IIb (bottom panel) samples at a similar epoch. SN~2017ivv has a peculiar H$\alpha$ P-Cygni profile compared with the fast-declining objects. While ASASSN-15oz, SN~2013ej and SN~2014G show a small absorption component together with the Cachito line \citep{Gutierrez17a}, SN~2017ivv and SN~2013by  display a flat absorption component. Although they have similar absorption profiles, the emission components are different. For SN~2013by the emission is broader and symmetric compared to SN~2017ivv. The slow-declining SN~II~2015bs, also included in this comparison, shows many similarities with SN~2017ivv: they only display hydrogen lines and \ion{He}{I}/\ion{Na}{I} D, the H$\alpha$ profile shows an asymmetric emission and flat absorption, while  \ion{Fe}{II} $\lambda5169$ starts to be visible. 
From the comparison with the SNe~IIb, the boxy H$\alpha$ profile observed in SN~2017ivv is similar to that of SN~1993J. This profile shape has been attributed to ejecta asymmetry, two line-forming regions for hydrogen, or CSM interaction \citep[e.g.,][]{Baron95,Hachinger12}. The other objects have a H$\alpha$ emission, with some signs of helium. In general, the bluer part of the spectrum of SN~2017ivv (below 5000\AA) shows considerably fewer lines than fast-declining SNe~II and SNe~IIb. This suggests that SN~2017ivv has either a higher temperature than the other SNe or the appearance of the lines is a metallicity effect \citep{Dessart14,Anderson16,Gutierrez18}. We note that the latter alternative is consistent to the low-metallicity of the host (see Section~\ref{galaxy}).

\subsection{Late phases}
\label{Snebular}

\begin{figure*}
\centering
\includegraphics[width=0.9\textwidth]{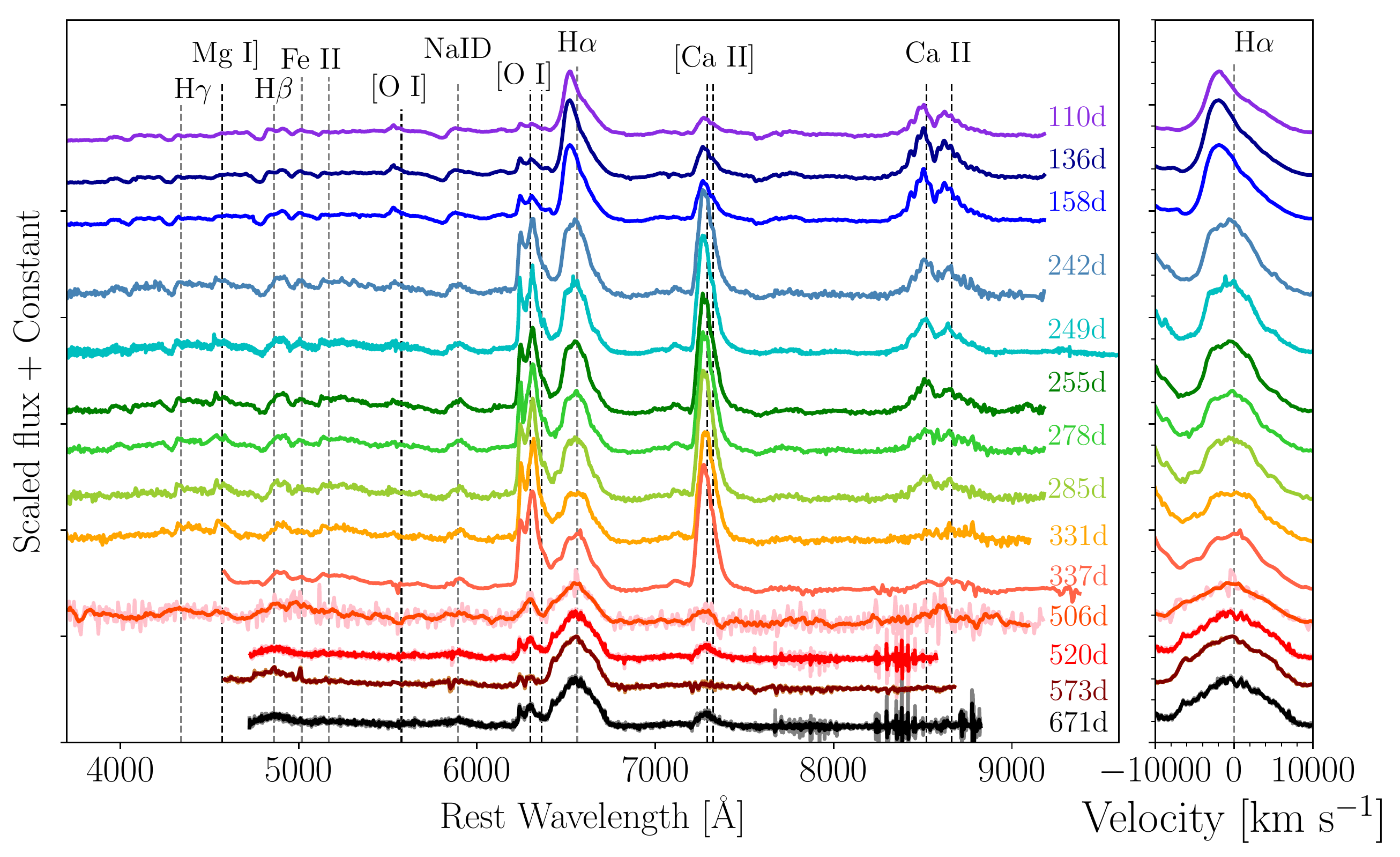}
\caption{\textbf{Left:} Optical spectra of SN~2017ivv ranging from 110 to 671 days from explosion. The phases are labeled on the right. Each spectrum has been corrected for
Milky Way reddening and shifted by an arbitrary amount for presentation. \textbf{Right:} Zoom around the H$\alpha$
P-Cygni profile in velocity space.}
\label{neb}
\end{figure*}

In the nebular phase, we obtained 14 spectra of SN~2017ivv between 110 and 671 days (Figure~\ref{neb}). During this phase, the spectra are mainly dominated by forbidden emission lines formed in low density regions. At 110 days, H$\alpha$ and the \ion{Ca}{II} NIR triplet are the strongest features in the spectrum. The forbidden lines of \ion{[O}{I]} $\lambda5577$, \ion{[O}{I]}  $\lambda\lambda6300$, 6364 and \ion{[Ca}{II]} $\lambda\lambda7291$, 7324 are also detected, but a slightly weaker. At bluer wavelengths, the iron absorption lines are still dominating the spectrum. The \ion{Na}{I} D doublet displays a clear residual absorption component. This surviving P-Cygni profile is visible until $\sim250$ days. Between 110 and 158 days, the spectra show only minor evolution, with the emission lines increasing gradually in strength relative to the continuum. 

From 158 to 242 days, the strength of the emission lines completely changes. At 158 days, H$\alpha$ is the strongest line in the spectrum, followed by the \ion{Ca}{II} NIR triplet and \ion{[Ca} {II]} $\lambda\lambda7291$, 7324. However, after 242 days, \ion{[Ca}{II]} becomes the strongest feature. The intensity of H$\alpha$ decreases while \ion{[O}{I]} $\lambda\lambda6300$, 6364 grows. At 242 days, \ion{[O}{I]} $\lambda5577$ emission is barely detected, \ion{Mg}{I]} and \ion{[Fe}{II]} $\lambda\lambda7155$, 7172 start to be detectable and the \ion{Ca}{II} NIR triplet diminishes. We also notice that H$\alpha$ develops a ``flat-topped" emission profile (right panel of Figure~\ref{neb}). H$\alpha$ starts to have a ``bridge'' with \ion{[O}{I]} $\lambda\lambda6300$, 6364. This ``bridge'' has been seen before in both SNe~IIL  (e.g. 2013by; \citealt{Black17} and 2014G; \citealt{Terreran16}) and SNe~IIb \citep[e.g. SN~1993J;][]{Matheson00}. From 337 to 671 days, the strength of \ion{[Ca} {II]} and \ion{[O}{I]} lines decrease dramatically, while the H$\alpha$ emission line becomes much broader; H$\alpha$ is the dominant feature and the ``bridge'' with \ion{[O}{I]} disappears. More details about the evolution of the nebular lines in SN~2017ivv and their connection with the SN~IIb class can be found in Section~\ref{Sneblines}.

\begin{figure*}
\centering
\includegraphics[width=15cm]{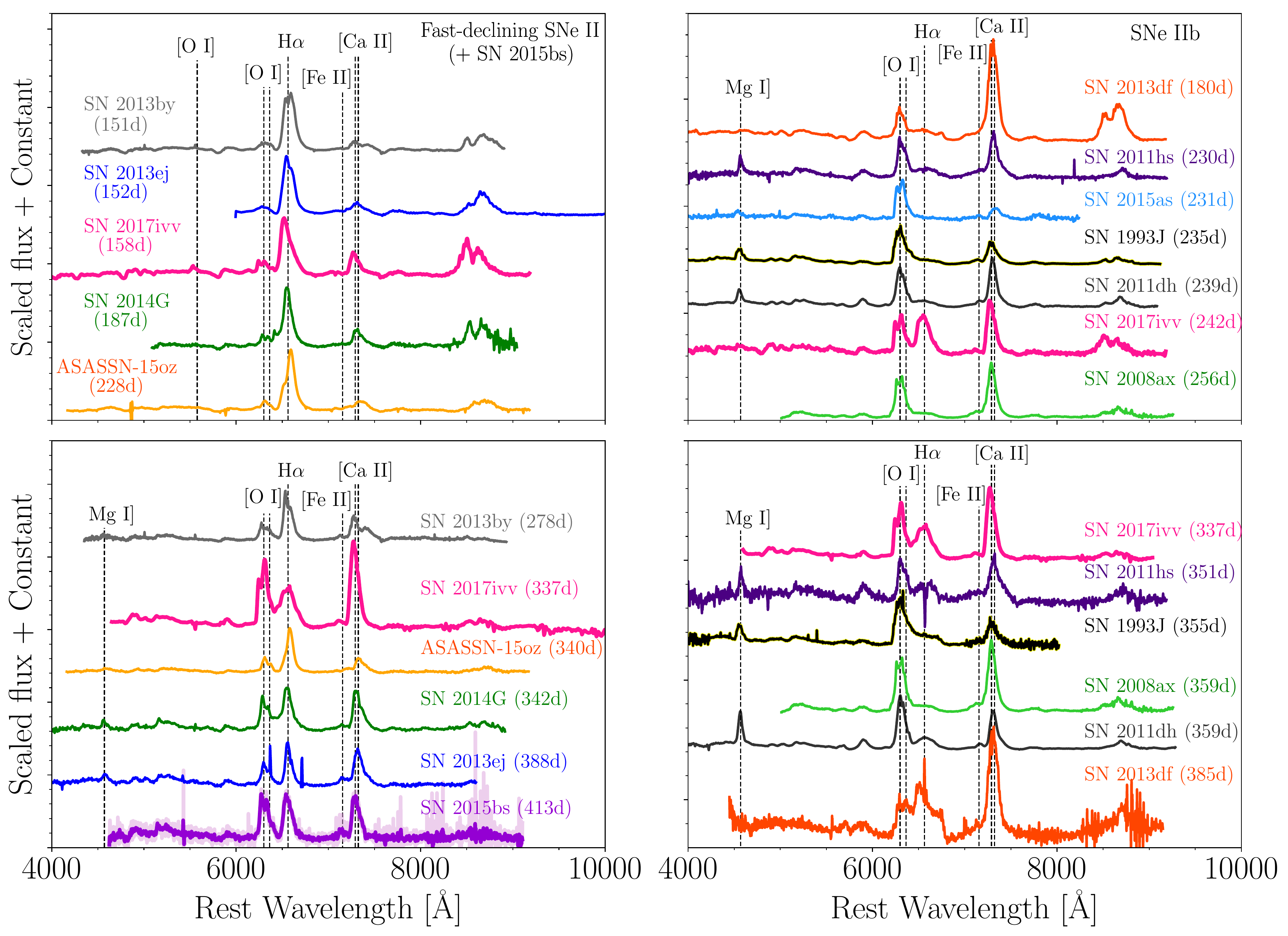}
\caption{Spectral comparison of SN~2017ivv in nebular phases. \textbf{Left panels:} Comparisons at $\sim158$ (top panel) and $\sim337$ days (bottom panel) with the fast-declining SN~II sample and SN~2015bs at similar epochs. Spectra are scaled to match H$\alpha$ and shifted for presentation. \textbf{Right panels:} Comparisons at $\sim242$ (top panel) and $\sim337$ days (bottom panel) with the SN IIb sample at similar epochs. Spectra are scaled to match \ion{[O}{I]} and shifted for presentation. All spectra have been corrected by Milky Way reddening and redshift.}  
\label{compneb}
\end{figure*}

A comparison of the nebular spectrum of SN~2017ivv at 158 and 337 days with spectra for the fast-declining sample and SN~2015bs at similar epochs is presented in Figure~\ref{compneb} (left panels). At around 158 days (top panel), all SNe show similar spectra, with H$\alpha$ being the dominant feature, followed by the \ion{Ca}{II} NIR triplet. One obvious feature observed in SN~2017ivv and missing in the other SNe is the blue-shifted of the emission line peaks. Blue-shifted lines can indicate asymmetries in the ejecta or radiative transfer effects (see more details in Section~\ref{Sneblines}). At around 338 days (bottom panel), SN~2017ivv has unusual strengths in its emission lines. While H$\alpha$ is the dominant feature in the fast-declining objects, in SN~2017ivv H$\alpha$ is weak and \ion{[Ca}{II]} is the strongest line. For the slow-declining SN~II 2015bs, the strength of H$\alpha$, \ion{[Ca}{II]} and \ion{[O}{I]} seem to be similar (see Section~\ref{LLH}). 
Another peculiar characteristic of SN~2017ivv is the strength of the \ion{[O}{I]} $\lambda\lambda6300$, 6364 doublet. According to \citet{Jerkstrand14}, \ion{[O}{I]} $\lambda6364$ is always equivalent (thick limit) or weaker (thin limit) than \ion{[O}{I]} $\lambda6300$. In SN~2017ivv, however, the relative strengths of the \ion{[O}{I]} doublet lines appear inverted. More details about this line is presented in Section~\ref{soxygen}.

In Figure~\ref{compneb} (right panels), SN~2017ivv is compared with the SN~IIb sample at around 242 and 337 days. At $\sim242$ days (top panel), all SNe show the \ion{[O}{I]} doublet feature with a blue-shifted emission peak. This blue-shift disappears in SN~2011hs and SN~2013df 100 days later (bottom panel). \ion{[Ca}{II]} is blue-shifted in SN~1993J, SN~2008ax and SN~2017ivv, red-shifted in SN~2015as, and at the rest in SN~2011hs and SN~2011dh. At 337 days, the only object with \ion{[Ca}{II]} still blue-shifted is SN~2017ivv. Overall, two evident differences between SN~2017ivv and the sample of SN~IIb are the strength of \ion{Mg}{I]}, which is weaker in the spectra of SN~2017ivv, and the strong and broad H$\alpha$ emission profile in SN~2017ivv. 

A broad H$\alpha$ emission profile has been identified in the late-time spectra of SNe~IIb (e.g. SN~1993J, SN~2008ax, SN~2011dh, SN~2011hs, SN~2013dh; see bottom panel) and has been linked to CSM interaction \citep[e.g.][]{Patat95, Matheson00, Taubenberger11, Maeda15}. Alternatively, \citet{Maurer10} explain the presence of H$\alpha$ as the result of mixed and strongly clumped H and He, while \citetalias{Jerkstrand15} show that in
radioactivity-powered models emission here is completely dominated by \ion{[N}{II]} $\lambda\lambda6548, 6583$, with only a minor H$\alpha$ contribution (see also \citealt{Fang18}). One should in general differentiate the early and later nebular evolution; several IIb SNe show consistency with [N II] in an early radioactivity-powered phase, transitioning to interaction and an H$\alpha$- dominated spectrum at $\gtrsim$500-700d. Although H$\alpha$ has been identified previously in many SNe~IIb, the strength and evolution observed in SN~2017ivv are somewhat unprecedented.

\subsection{Expansion velocities}
\label{Svel}

\begin{figure}
\centering
\includegraphics[width=\columnwidth]{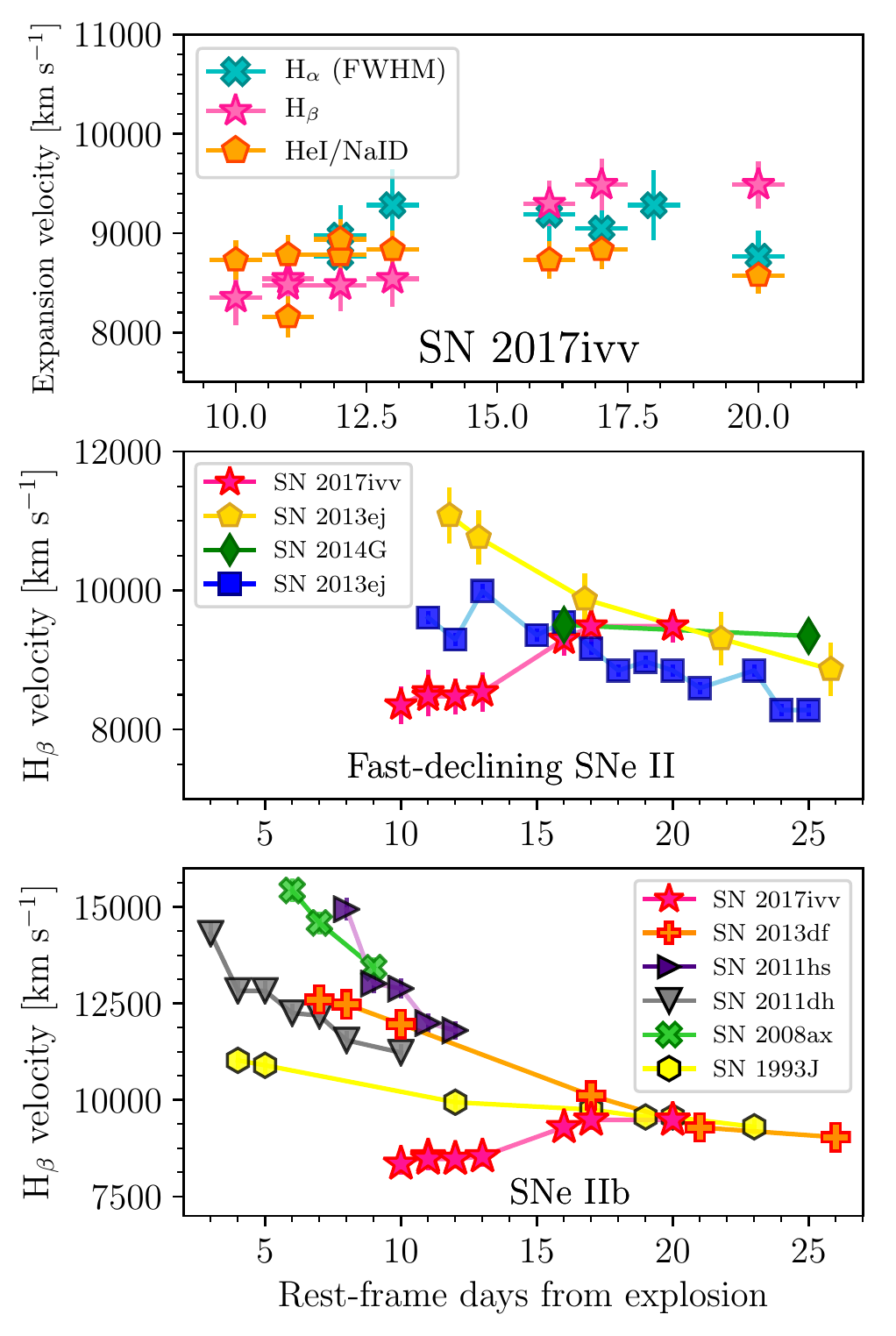}
\caption{\textbf{Top panel:} Expansion velocity evolution of SN~2017ivv over the first 20 days. The H$\alpha$ velocity was obtained from the FWHM of the emission line, while H$\beta$ and \ion{He}{i} were derived from the minimum of the absorption line. \textbf{Middle panel:} Comparison of the H$\beta$ velocity of SN~2017ivv with the fast-declining sample. \textbf{Bottom panel:} Comparison of the H$\beta$ velocity of SN~2017ivv with SN~IIb sample.}
\label{velp}
\end{figure}

The broad lines in the early spectra of SN~2017ivv indicate large expansion velocities. In Figure~\ref{velp} (top panel), we present the velocity evolution of H$\alpha$, H$\beta$ and \ion{He}{i} $\lambda5876$ during the first 20 days. These velocities were mostly obtained from the minimum flux of the absorption. In the case of  H$\alpha$, it was derived from the full width at half maximum (FWHM) of the emission component (see \citealt{Gutierrez17a}). All lines have nearly constant velocities, although the H$\beta$ velocity seems to increase by around $1000$ km s$^{-1}$ from 10 to $\sim15$ days. \citet{Folatelli14a} found similar velocity evolution for SN~2010as and some transitional Type Ib/c SNe, the so-called ``flat-velocity Type IIb Supernovae". The behaviour was most notable for the helium lines, but in some objects it was a general feature of the photosphere, and could be evidence of the the presence of a dense shell inside the SN ejecta. 

The time evolution of the expansion velocity of H$\beta$ in SN~2017ivv is compared with the fast-declining and SN~IIb samples in the middle and bottom panels of Figure~\ref{velp}, respectively.  
The velocity evolution of SN~2017ivv during the first 16 days is different from all other SNe because they all have rapidly evolving velocities. After 16 days, the fast-declining, SN~2014G, and the SNe~IIb, 1993J and 2013df have similar velocities than SN~2017ivv. 

\begin{figure}
\centering
\includegraphics[width=\columnwidth]{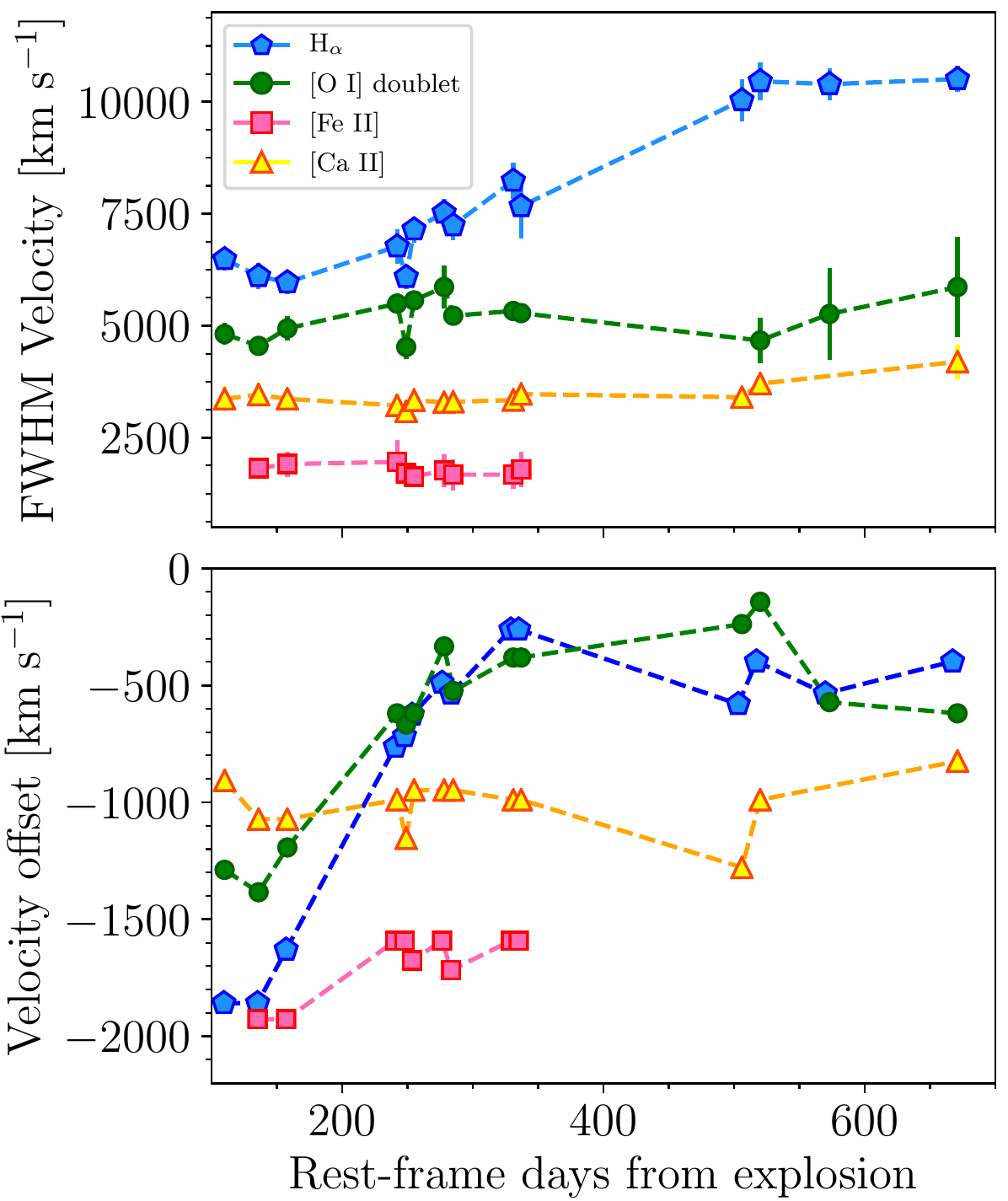}
\caption{\textbf{Top panel:} FWHM velocity of H$\alpha$ (orange), \ion{[O}{I]} (blue), \ion{[Ca}{II]} (magenta) and \ion{[Fe}{II]} (green) emission lines of SN~2017ivv. \textbf{Bottom panel:} Velocity offset of emission lines in SN 2017ivv. Same colours and markers as in the top panel.}
\label{velocityn}
\end{figure}

During the nebular phase, the velocity of the emission lines is measured using the line FWHM. The FWHM velocities of H$\alpha$, \ion{[O}{I]} $\lambda\lambda6300$, 6364, \ion{[Ca}{II]} $\lambda\lambda7291$, 7324 and \ion{[Fe}{II]} $\lambda\lambda7155$, 7172 are measured by fitting Gaussian profiles to each line. When a line is affected by another one, we first remove the potential contribution of the latter and then we fit the line. Examples of Gaussian fits at four different epochs are presented in Figure~\ref{gfits}. Finally, all FWHM were corrected for the resolution of the instrument following the prescriptions of \citet{Maguire12}. The FWHM velocities are reported in Table~\ref{measurements} and shown in the top panel of Figure~\ref{velocityn}. Here, we see that \ion{[O}{I]}, \ion{[Ca}{II]} and \ion{[Fe}{II]} show little velocity evolution. After 337 days, \ion{[Fe}{II]} is barely detected, so no estimation of its velocity is presented. Flat FWHM evolution in SNe~II has been reported before by \citet{Maguire12,Silverman17} although only after $\sim300$ days. Before 300 days, decreasing velocities were measured.

SN~2017ivv has an H$\alpha$ FWHM velocity increasing with time from FWHM$=5970$ km s$^{-1}$ at 158 days to FWHM$=10510$ km s$^{-1}$ at 671 days. The increase in the FWHM of H$\alpha$ is large (more than 4000 km s$^{-1}$ in $\sim510$ days). For \ion{[O}{I]} $\lambda\lambda6300$, 6364, the FWHM velocity remains at $\sim5000$ km s$^{-1}$. This FWHM value is similar to that found by \citet{Taubenberger09} for stripped-envelope SNe, and larger than the average value for SNe~II (3000 km s$^{-1}$; \citealt{Maguire12}). 

On the bottom panel of Figure~\ref{velocityn}, we present the velocity offset of the emission lines. All of the lines are blue-shifted with velocity offsets between $-2000$ and $-400$ km s$^{-1}$. The offset slowly decreases with time, but without reaching zero velocity. Generally, for opacity caused by atoms (in contrast to dust) the blue-shifts of emission peaks gradually disappear as the optical depths decrease with time \citepalias[e.g.][]{Jerkstrand15}. There is a weak such trend up to 300d for H$\alpha$, [Ca II], [Fe II] and the [O I] doublet taken as a whole. However, after that time the blue-shifts are retained at roughly constant values. Such levels of blue-shifts observed in SN~2017ivv have been seen before in SN~1993J and were explained by an asymmetric mass distribution for the emitting material in the envelope \citep{Spyromilio94}. More details can be found in Section~\ref{Sneblines}).

\section{Analysis and discussion}
\label{analysis}

\subsection{A low-metallicity host and its  effects on the SN spectra}
\label{LLH}

\begin{figure}
\centering
\includegraphics[width=\columnwidth]{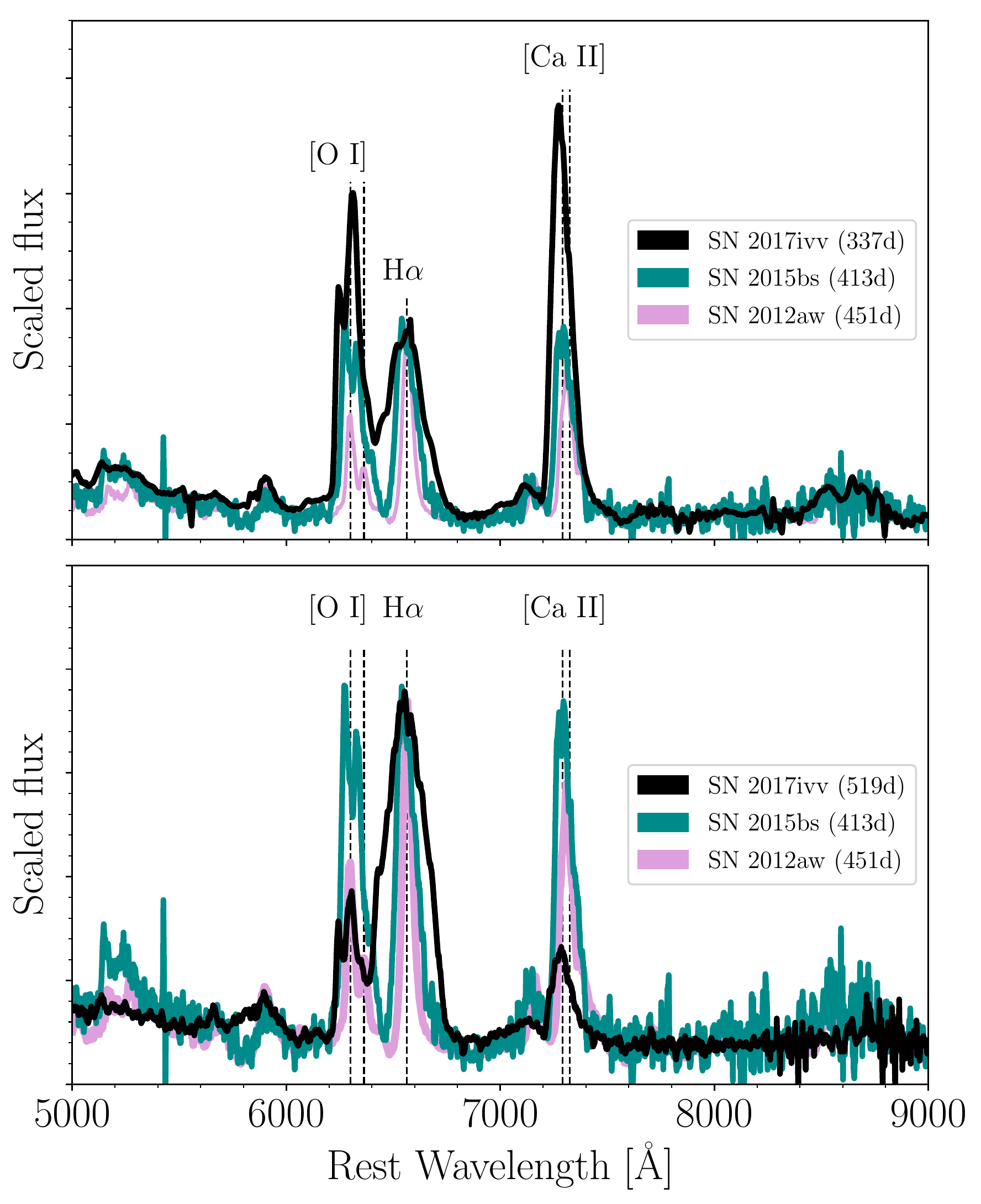}
\caption{Spectral comparison of SN~2017ivv at 337 (top) and 519 days (bottom) with SN~2015bs at 413 days. For reference, we use the spectrum of the slow-declining SN~II, SN~2012aw at 451 days \citep{Jerkstrand14}. Spectra are scaled to match the H$\alpha$ line.}   
\label{neb15bs}
\end{figure}

The host galaxy of SN~2017ivv has the lowest luminosity (M$_r\sim-10.3$ mag, see Section~\ref{galaxy}) of any SN~II host found to date. The previous record, the host of SN~2015bs was about two magnitudes more luminous \citep{Anderson18}.

\citet{Dessart13,Dessart14} presented SN~II spectral models produced from progenitors with metallicities of 0.1, 0.4, 1, and 2 times solar metallicity (\Zsun). \citet{Dessart14} predicted a dependence of the strength of the metal lines observed during the recombination phase with the progenitor metallicity. \citet{Anderson16} tested this prediction using more than 100 SNe~II, and found a trend between the strength of \ion{Fe}{II} $\lambda5018$ \AA\ and the local metallicity (oxygen abundances). Similar results were found by \citet{Taddia16} and \citet{Gutierrez18}. 

\citet{Anderson18} found a remarkable similarity between SN~2015bs at 57 days and the 0.1 \Zsun\ model at 50 days. Unfortunately, for SN~2017ivv this analysis is impossible to make due to the lack of observations between 20 and 100 days. However, the effects of metallicity could be also be explored during the nebular phase. Through comparison to the nebular-phase spectral models, \citet{Anderson18} argued that SN~2015bs has a high mass progenitor since the strength of the \ion{[O}{I]} $\lambda\lambda6300$, 6364 was bounded by the 15 \Msun\ and 25 \Msun\ progenitor models. This result was based on the relation between the helium-core and ZAMS mass \citep{Woosley95}.

We compared the nebular spectra of SN~2017ivv and SN~2015bs in Figure~\ref{neb15bs}. Because we do not have observations at similar epochs, we use the nebular spectrum of SN~2015bs at 413 days, and two spectra of SN~2017ivv at 337 days (top panel) and 519 days (bottom panel). In the top panel we see that the strength of the \ion{[O}{i]} line is larger in SN  2017ivv, with the redder peak stronger than the bluer (more details can be found in Section~\ref{soxygen}). 
In SN~2015bs, H$\alpha$ has a narrower profile, while \ion{Ca}{II} is weaker. As the \ion{[O}{I]} doublet is a good tracer of the ZAMS progenitor mass, this comparison could imply a more massive progenitor for SN~2017ivv. However, a higher helium core mass (strong \ion{[O}{I]}) can be also related to metallicity. \citet{Woosley02} predict that for massive stars the helium-core mass increases as metallicity decreases. As SN~2017ivv exploded in a metal-poor environment, this possibility cannot be ruled out.

When SN~2017ivv is compared at 519 days with SN~2015bs (bottom panel), the similarities disappear. In SN~2017ivv all emission lines are less prominent than those of SN~2015bs. The changes in the emission lines and the broad profile of H$\alpha$ could indicate an interaction between the ejecta and the CSM (see Section~\ref{Sneblines}). This hypothesis is also supported by the shallow slope of the light curves at late times.

\subsection{Nebular line evolution}
\label{Sneblines}

\begin{figure}
\centering
\includegraphics[width=\columnwidth]{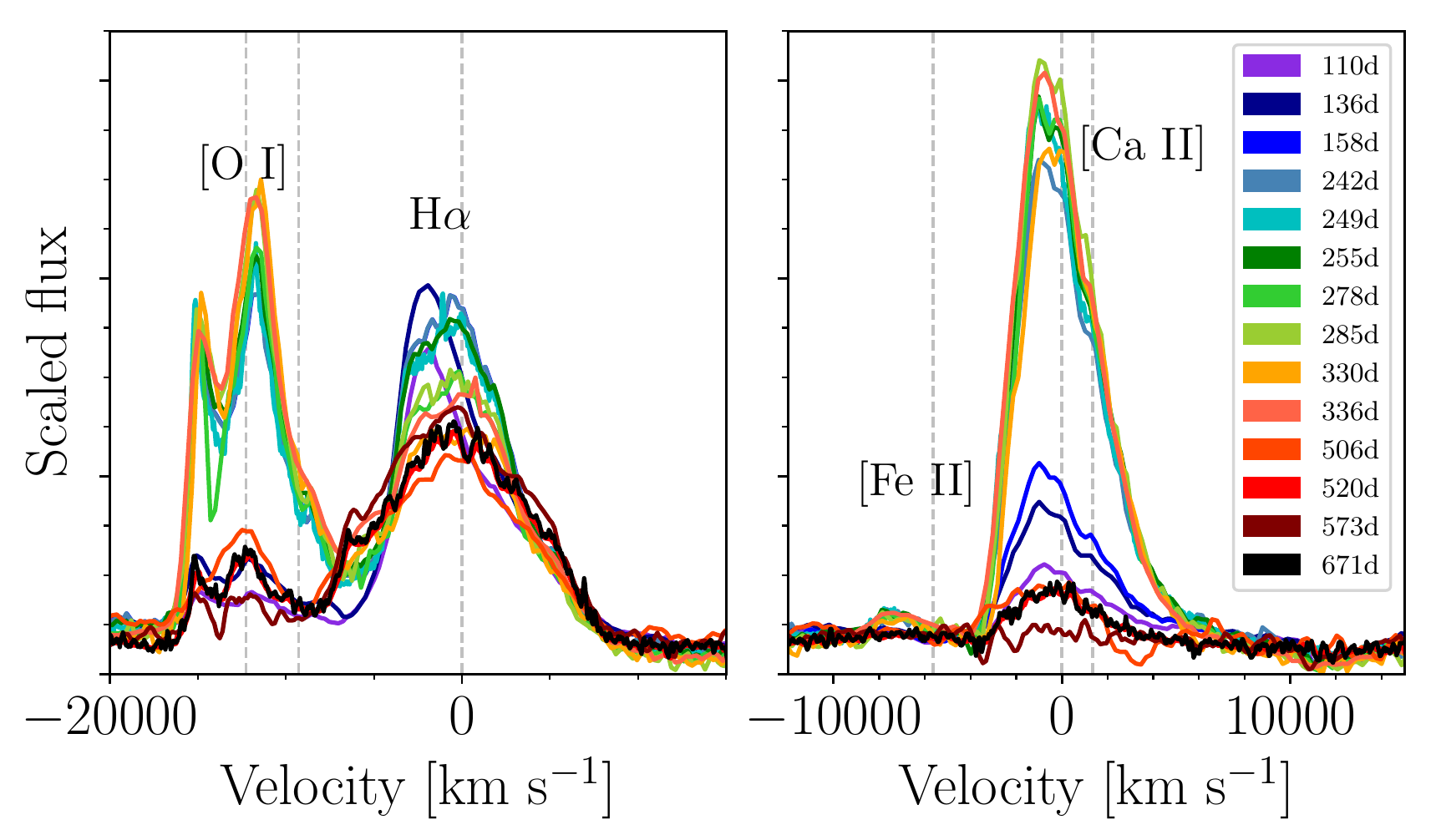}
\caption{Observed profiles of SN~2017ivv in the nebular phase. \textbf{Left panel:} \ion{[O}{I]} $\lambda\lambda6300$, 6364 and H$\alpha$ emission lines. \textbf{Right panel:} \ion{[Fe}{II]} $\lambda\lambda7155$, 7172 and \ion{[Ca}{II]} $\lambda\lambda7291$, 7324 emission lines. The phases are labeled on the right.}  
\label{nebevol}
\end{figure}

One of the most interesting characteristics of SN~2017ivv is the unusual late-time spectral evolution. Figure~\ref{nebevol} shows the profiles of the H$\alpha$, \ion{[O}{I]}, \ion{[Fe}{II]} and \ion{[Ca}{II]} lines between 110 and 671 days. Details of the evolution of each line are presented below.

\subsubsection{H$\alpha$ emission line profile}
\label{sec:Ha}

The evolution of the H$\alpha$ emission line is shown in the left panel of Figure~\ref{nebevol}. Overall, H$\alpha$ is strongly asymmetric with a slowly dropping blue-shift in the emission peak. The velocity offset (shown in the bottom panel of Figure~\ref{velocityn}) declines from $\sim-1860$ km s$^{-1}$ at 110 days to $\sim-400$ km s$^{-1}$ at 671 days. Blue-shifted lines during the nebular phase have been related to asymmetries in the ejecta \citep[e.g.][]{Utrobin09} or radiative transfer effects due to dust \citep[e.g.][]{Sahu06, Maguire10, Maguire12} or line opacities \citep{Jerkstrand15}.
There are negligible changes in the intensity and width of the  H$\alpha$ line between 110 to 158 days. After 158 days, the H$\alpha$ flux drops significantly (see Figure~\ref{flux}), but the line is getting broader. Specifically, from 158 to 242 days, the H$\alpha$ flux declines around 20\%. Although the H$\alpha$ flux is predicted to decline at late-times \citep[e.g.,][]{Chevalier94}, observations of SNe~II with well-sampled nebular coverage \citep[e.g SN~2004et,][]{Sahu06} have shown that this drop usually happens gradually after $\sim300-400$ days. Similar decreases in the H$\alpha$ fluxes have been seen for a few  SNe~II (e.g. SN~1980K; \citealt{Fesen99}), but only after a few years. 

After 242 days, H$\alpha$ develops a ``flat-topped" emission profile, has a weaker strength compared with \ion{[O}{I]} and \ion{[Ca}{II]}, and displays a different profile from that observed in SNe~II at similar phases (Figure~\ref{compneb}). A and flat-topped H$\alpha$ emission can be associated with CSM interaction in a roughly spherical shell \citep{Chevalier85, Jeffery90, Chevalier94}, however, for SN~2017ivv, the fade and shape of H$\alpha$ seem to be related to the end of the hydrogen recombination.
At 242 days, the broad H$\alpha$ emission is becoming blended with \ion{[O}{I]} $\lambda\lambda6300$, 6364. The ``bridge'' between these two lines grows for more than 200 days. In the spectrum at 506 days, the ``bridge'' begins to recede, while the H$\alpha$ emission develops a blue shoulder, that makes it even broader.

\begin{figure}
\centering
\includegraphics[width=\columnwidth]{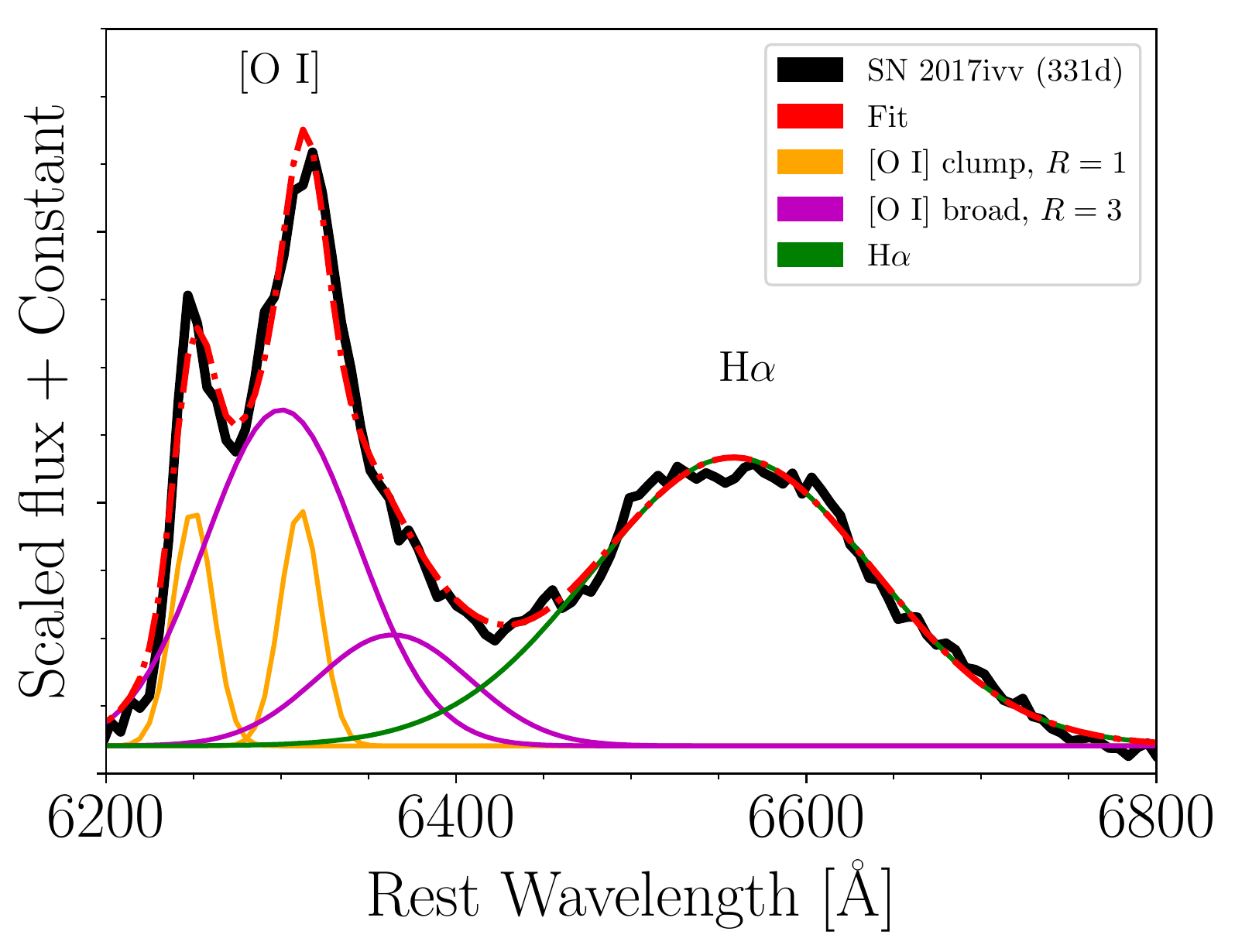}
\caption{Best fit of the 6200 -- 6800 \AA\ region of the spectrum of SN~2017ivv at 331 days. The observed spectrum is shown in black, the overall fit in red, the individual optically thin and optically thick components for \ion{[O}{I]} in purple and orange, and the individual Gaussian profile for H$\alpha$ in green.}   
\label{nebfit}
\end{figure}

\citet{Terreran16} showed that the ``bridge'' between H$\alpha$ and \ion{[O}{I]} observed in the SN~2014G can be explained in terms of boxy line profiles due to CSM interaction. Similar line structures are seen in SNe~IIb such as SN~1993J, SN~2008ax, SN~2011fu, SN~2011hs and SN~2013df.
For SN~2017ivv, we reproduce the emission profiles in the 6200 -- 6800 \AA\ region and the best fit is presented in Figure~\ref{nebfit}. 
To fit this region simultaneously, we considered a two-component model for the \ion{[O}{I]} doublet and a single Gaussian for H$\alpha$. Therefore, the profile fit consists of six parts: a linear continuum; two narrow Gaussians with the same FWHM ($\sim$1300 \kms) and strength, separated by 64 \AA, and with a blue-shift of $\sim2400$ \kms with respect to their rest-frame wavelengths (orange); two broad Gaussians centred at 6300 and 6364, which have the same FWHM ($\sim$4900 \kms) but the component at 6364 has a third the strength of the component at 6300; and a single (FWHM$\sim$8700 \kms) Gaussian centered at $6563\pm10$ \AA\ (green).
Here, the blue-shifted and narrow profiles in \ion{[O}{I]} could be interpreted as a pseudo-spherical distribution of the oxygen-rich ejecta with a localized clump travelling towards the observer with a projected blue-shifted velocity.
As see in the figure, the bridge between H$\alpha$ and \ion{[O}{I]} can be reproduced by the \ion{[O}{I]} $\lambda6364$ component.

Trying to reproduce the nebular spectra of SN~1993J, SN~2008ax, and SN~2011dh,   \citetalias{Jerkstrand15} found that \ion{[N}{II]} $\lambda\lambda6548$, 6583 emission was more important than H$\alpha$. In their models, the emission near 6550 \AA\ can be caused by \ion{[N}{II]} $\lambda\lambda6548$, 6583 because the hydrogen envelopes in SNe~IIb are too small to produce any detectable H$\alpha$ after 150 days. At the same time, strong \ion{[N}{II]} emission follows naturally from the quite massive He/N shell ($\sim 0.5-1~$\Msun), in which N is a much more efficient coolant than He. \citet{Fang18}, analysing seven SNe~IIb and two SN~Ib, found support for this interpretation.
Although the \ion{[N}{II]} $\lambda\lambda6548$, 6583 is a possibility, we suggest the emission at $\sim6550$ \AA\ is related to H$\alpha$. This suggestion is supported by the  fact that H$\alpha$ is stronger than the predicted \ion{[N}{II]} emission. After 500 days, the broad H$\alpha$ emission indicates that the line arises from both the radioactive heating of hydrogen-rich ejecta from interior regions and the reverse shock interacting with high-velocity outer hydrogen-rich material.

\subsubsection{Oxygen emission lines}
\label{soxygen}

During the nebular phase, CC-SNe exhibit oxygen emission lines from \ion{O}{I} $\lambda7774$, \ion{[O}{I]} $\lambda5577$, and \ion{[O}{I]} $\lambda\lambda6300$, 6364. The \ion{O}{I} $\lambda7774$ line is not detected in the spectra of SN~2017ivv. The \ion{[O}{I]} $\lambda5577$ and \ion{[O}{I]} $\lambda\lambda6300$, 6364 lines become visible at the beginning of the nebular phase, but \ion{[O}{I]} $\lambda5577$ disappears more rapidly due to its higher temperature sensitivity. 
In SN~2017ivv, \ion{[O}{I]} $\lambda5577$ fades with time and is barely detected after 244 days. The \ion{[O}{I]} $\lambda\lambda6300$, 6364 profile is detected during the nebular phase showing a double-peaked structure. The origin of the double-peaked \ion{[O}{I]} $\lambda\lambda6300$, 6364 emission line has been discussed by several authors \citep[e.g.,][and references therein]{Milisavljevic10}. A double peaked structure can be explained by a torus or disk-like geometry. \citep[e.g.][]{Mazzali01, Maeda02, Mazzali05, Maeda06, Maeda08, Modjaz08, Taubenberger09}. An alternative explanation, based on the separation of the two peaks ($\sim64$ \AA), suggests that the these peaks may originate from the two lines of the doublet from a single emitting source on the front of the SN ejecta which move towards the observer \citep{Milisavljevic10}. \citet{Maurer10} suggested that the double-peaked oxygen emission can also be caused by high-velocity (above $\sim11000$ km s$^{-1}$) H$\alpha$ absorption.

For SN~2017ivv, just the first scenario is fully plausible. Like H$\alpha$, both \ion{[O}{I]} $\lambda5577$ and \ion{[O}{I]} $\lambda\lambda6300$, 6364 show blue-shifted emission profiles over the entire nebular phase. For \ion{[O}{I]} $\lambda5577$, the blue-shift of the peak decreases from $\sim-2500$ at 110 days to $\sim-1900$ km s$^{-1}$ at 242 days. The velocity offset of \ion{[O}{I]} $\lambda\lambda6300$, 6364 line goes down from $\sim-1290$ at 111 days to $\sim-145$ km s$^{-1}$ at 520 days (bottom panel of Figure~\ref{velocityn}). As we see, these emission lines keep their blue-shifted velocities during all nebular observations. Similar behaviour was reported by \citet{Milisavljevic10} for stripped-envelope SNe with double-peaked  \ion{[O}{I]} profiles.  
Multiple scenarios have been proposed to explain the blue-shift of \ion{[O}{I]} in these objects, including ejecta geometry, contamination from other emission lines and opacity in the core of the ejecta \citep[for more details see][and references therein]{Taubenberger09}. Based on the blue-shifts observed in all emission lines of SN 2017ivv, and their slow drop with time, the most consistent interpretation is asymmetries in the ejecta (geometric scenario; see Section~\ref{smodels}). In the early nebular phases, indications of additional blue-shifting, likely due to residual atomic opacity, are also seen.

\begin{figure}
\centering
\includegraphics[width=\columnwidth]{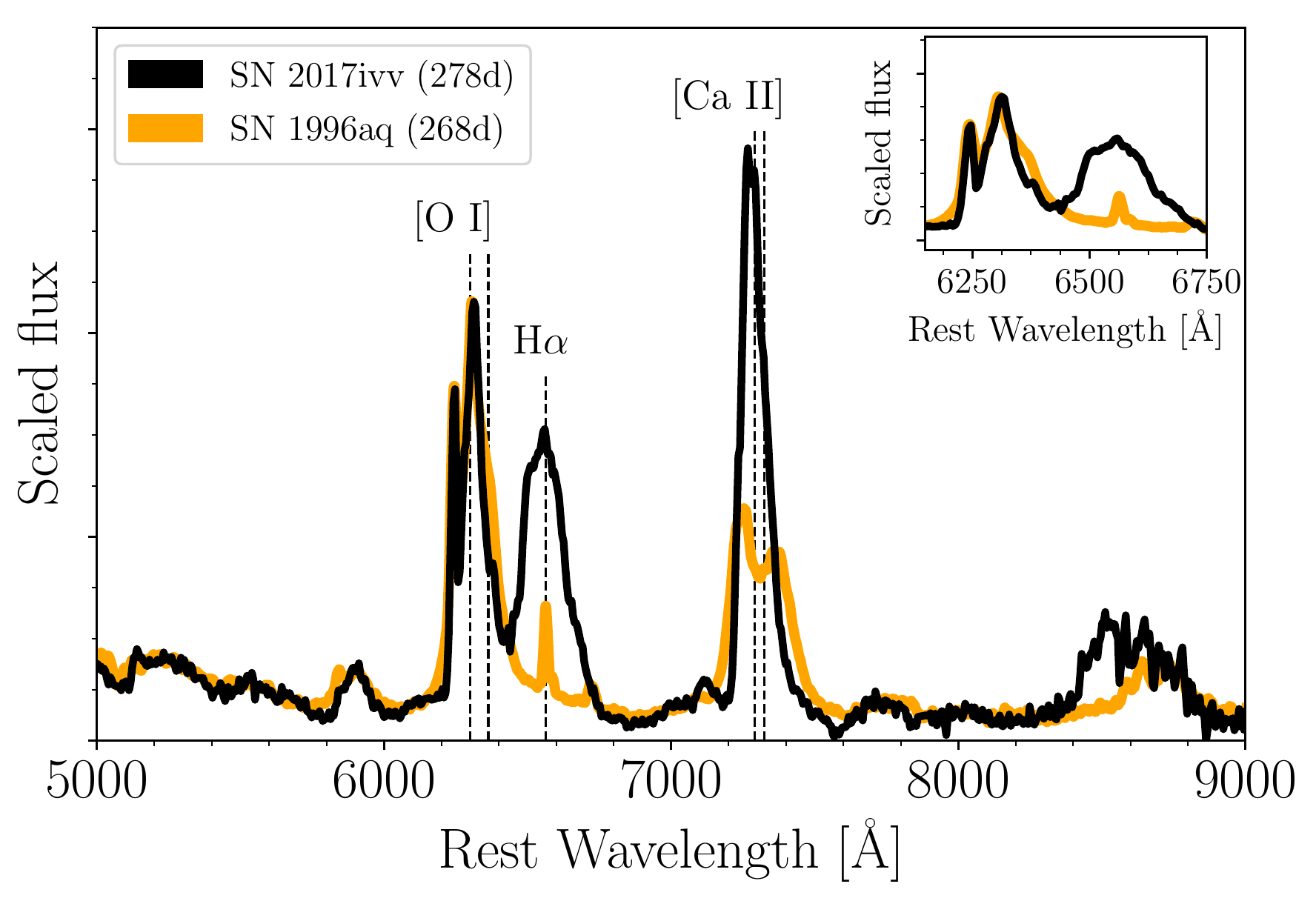}
\caption{Spectral comparison of SN~2017ivv at 278d with the type Ib SN~1996aq at 268 days. Spectra are scaled to match the  \ion{[O}{I]} $\lambda\lambda6300$, 6364 doublet.}   
\label{neb96aq}
\end{figure}

On the left panel of Figure~\ref{nebevol}, we show the evolution of the double-peaked \ion{[O}{I]} $\lambda\lambda6300$, 6364 profile of SN~2017ivv. The \ion{[O}{I]} line exhibits a quite unusual evolution. Between 110 and 158 days, \ion{[O}{I]} shows two emission peaks that resemble the nebular profiles of SN~II. During this period, the blue and red peaks have equal strengths, expected in the optically thick limit \citep{Jerkstrand14}. In the optically thin limit, \ion{[O}{I]} $\lambda6364$ is always three times weaker than \ion{[O}{I]} $\lambda6300$ \citep[e.g.][]{Spyromilio91,Li92,Taubenberger09,Jerkstrand14}, but in SN~2017ivv, from 242 to 337 days, the redder peak gradually becomes stronger than the bluer one. Some stripped-envelope SNe (e.g. SN~1996aq \citealt{Valenti08,Taubenberger09} and SN~2003jd; \citealt{Mazzali05, Valenti08}) have shown this behaviour. In Figure~\ref{neb96aq}, we compare SN~2017ivv at 278 days with one of these objects, SN~1996aq \citep{Taubenberger09} at 268 days. The \ion{[O}{I]} profile in both objects is quite similar: identical strengths in the red and blue peaks together with comparable line width. 

\citet{Maeda02} showed that the emission profiles in SNe~Ic can be affected by an aspherical explosion, and the double-peaked profiles  can be produced when the degree of asymmetry is high. Similar results were also found by \citet{Mazzali05}. Using an aspherical explosion model, they found that the double-peaked profile of the \ion{[O}{I]} line in SN~2003jd can be reproduced by a jet axis oriented $>70^{\circ}$ away from the line of sight. Asymmetry with respect to the equatorial plane of the ejecta are shown to be able to give different strengths of the two peaks; for fixed viewing angle with respect to $z$ (the axis of the bipolar explosion asymmetry), the red or blue peak can be stronger depending on whether angle is with respect to $+z$ or $-z$. Therefore, a geometric scenario fully supports the blue-shifted emission lines and the double-peaked structure in \ion{[O}{I]}.

While the two peaks of \ion{[O}{I]} separated by $\sim63-68$ \AA\ ( Figure~\ref{gfits} in the Appendix) support the doublet nature of the \ion{[O}{I]} line emission, the structure of the \ion{[O}{I]} profile could exclude this possibility. This is because the asymmetry of the lines (the redder peak stronger than the blue peak) can be simply explained by an aspherical explosion. Finally, we rule out the latter scenario (high-velocity H$\alpha$ absorption) based on the Gaussian analysis performed in Section~\ref{sec:Ha} (Figure~\ref{nebfit}).

After 506 days, the two emission peaks of \ion{[O}{I]} $\lambda\lambda6300$, 6364 show again equal strength, resembling the first three spectra of the nebular phase. The decreasing\ion{[O}{I]} intensity can be an effect of CSM interactions.

\subsubsection{Calcium emission lines}

The evolution of the \ion{[Ca}{II]} $\lambda\lambda7291$, 7324 emission line profile is shown in the right panel of Figure~\ref{nebevol}. During the nebular observations, the strength of \ion{[Ca}{II]} changes significantly. From 110 and 158 days, \ion{[Ca}{II]} slowly evolves, with a equivalent width (EW) from 161 \AA\ increasing to 418 \AA, and by 242 days it goes up to $\sim1200$ \AA. At 506 days, the pEW of the line decreases considerably to $\sim150$ \AA. The decrease on the strength of the lines after 500 days could again be related to CSM interaction.

A double-peaked structure is detected in the top of the \ion{[Ca}{II]} line between 110 and 337 days, becoming more evident from 242 to 337 days. In addition to this double-peaked structure, \ion{[Ca}{II]} shows a red shoulder which could be attributed to \ion{[Ni}{II]} $\lambda7378$ \citep{Jerkstrand15a}. In terms of velocity, both the FWHM velocity and velocity offset (Figure~\ref{velocityn}) show little evolution. The FWHM velocity is $\sim4000$ km s$^{-1}$, which lies in the observed range of both SN~II \citep{Maguire10,Silverman17} and SNe~IIb.

\subsection{Temperature estimation from the ratio of the oxygen lines}
\label{stemp}

The ejecta temperature at a given time can be constrained by the ratio of \ion{[O}{I]} $\lambda5577$ to \ion{[O}{I]} $\lambda\lambda6300$, 6364 \citep[e.g.,][]{Houck96,Jerkstrand14}. Both \ion{[O}{I]} lines are excited by thermal collisions, but as they have different excitation energies, their line ratio depends on the temperature \citep[e.g.][]{Fransson89, Jerkstrand14}. The \ion{[O}{I]} $\lambda5577$ line starts deviating from local thermodynamic equilibrium (LTE) after $\sim$250d in SNe~II and after $\sim$150d in SNe~IIb \citep{Jerkstrand14,Jerkstrand15}. At too early phases ($\lesssim 100$d) radiative transfer and line blending effects become problematic. An LTE line ratio analysis is therefore most suitable at $\sim$ 150-200d.
As shown in Figure~\ref{neb}, \ion{[O}{I]} $\lambda5577$ is clearly detected in our first four nebular spectra, from 110 to 242 days.

The temperature given a certain line ratio of the \ion{[O}{I]} lines, assuming LTE, is given by the following equation

\begin{center}
$ \frac{L_{5577}}{L_{6300,6364}}=38\times$ exp($\frac{-25790 K}{T}$)$\frac{\beta_{5577}}{\beta_{6300,6364}}$
\end{center}

{\noindent}where the $\beta_{5577}/\beta_{6300,6364}$ ratio ($\beta_{ratio}$) must be in the range 1--2 if partial transitioning to optical thinness for \ion{[O}{I]} $\lambda\lambda6300$, 6364 is either observed or can be inferred from any reasonable values of oxygen mass given the epoch and expansion velocities \citep{Jerkstrand14}. Assuming $\beta_{5577}/\beta_{6300,6364}=1$, 1.5 and 2, we find the temperatures listed Table~\ref{temp} (columns 4, 5 and 6, respectively). In this table, we also report the measured line luminosities of \ion{[O}{I]} $\lambda5577$ and \ion{[O}{I]} $\lambda\lambda6300$, 6364, and the estimated oxygen mass corresponding to each temperature (see Section~\ref{somass}).
Looking at the temperatures obtained for SN~2017ivv at around 158 days with those calculated for SN~1993J, SN~2008ax, and SN~2011dh at similar epochs \citepalias{Jerkstrand15}, we find that SN~2017ivv has a temperature comparable with these type IIb events. 
Unfortunately, \ion{[O}{I]} $\lambda5577$ is not clearly detected in the majority of late spectra of SNe~II, but some weak detections have been made before (e.g. SN~2012aw; \citealt{Jerkstrand14}), although at later phases. 

\begin{table*}
\centering
\footnotesize
\caption{Local thermodynamic equilibrium (LTE) temperatures and \ion{O}{I} masses estimated by using the \ion{[O}{I]} $\lambda5577$ to \ion{[O}{I]} $\lambda\lambda6300$, 6364 ratio.}
\label{temp}
\begin{tabular}[t]{cccccccccccc}
\hline
\hline
Time   &	L$_{5577}$            &  L$_{6300,6364}$         &   T$_{\beta_{ratio}=1}$ & T$_{\beta_{ratio}=1.5}$ & T$_{\beta_{ratio}=2}$ & M(\ion{O}{I})$_{\beta_{ratio=1}}^{\ast}$ & M(\ion{O}{I})$_{\beta_{ratio=1.5}}^{\ast}$ &  M(\ion{O}{I})$_{\beta_{ratio=2}}^{\ast}$\\
(Days) & (10$^{38})$ erg s$^{-1}$ & (10$^{38})$ erg s$^{-1}$ &            (K)          &        (K)              &            (K)        &    (\Msun)   &    (\Msun)     &    (\Msun)       \\
(1)    & (2)  &  (3)  &  (4)    &    (5)  &  (6)    &  (7) &  (8) & (9)   \\ 
\hline
\hline
110    & 9.51 & 25.01 & 5602 & 5149 & 4869 & 0.30 & 0.42 & 0.55  \\  
136    & 7.35 & 26.71 & 5233 & 4836 & 4588 & 0.42 & 0.60 & 0.78  \\  
158    & 6.04 & 30.90 & 4894 & 4544 & 4325 & 0.66 & 0.94 & 1.22  \\  
242    & 1.04 & 15.22 & 4078 & 3832 & 3675 & 0.82 & 1.18 & 1.52  \\  
\hline 
\end{tabular}
\begin{list}{}{}
\item Note -- Columns: (1) Time from explosion; (2) Line luminosity of  \ion{[O}{I]} $\lambda5577$; (3) Line luminosity of \ion{[O}{I]} $\lambda\lambda6300$, 6364; (4) Temperature assuming $\beta_{ratio=1}$; (5) Temperature assuming $\beta_{ratio=1.5}$; (6) Temperature assuming $\beta_{ratio=2}$; (7) \ion{[O}{I]} mass corresponding to the temperature in column 4; (8) \ion{[O}{I]} mass corresponding to the temperature in column 5; (7) \ion{[O}{I]} mass corresponding to the temperature in column 6.
\item $^{\ast}$ Assuming $\beta_{6300,6364}=0.5$.
\end{list}
\end{table*}

\subsection{Oxygen mass}
\label{somass}

Using the temperature estimated in Section~\ref{stemp}, the \ion{O}{I} mass can be derived from the oxygen luminosity as follows 

\begin{center}
M$_{O}=\frac{L_{6300,6364}/\beta_{6300,6364}}{9.7\times 10^{41} \mathrm{erg s^{-1}}} \times exp(\frac{22720 K}{T})$ \Msun 
\end{center}

{\noindent}\citep{Jerkstrand14}. Taking $\beta_{6300,6364}=0.5$, we obtain the \ion{O}{I} masses presented in columns 7, 8 and 9 of Table~\ref{temp}. 
Comparing the value at $\sim158$ days to those obtained for the SN~IIb 1993J, SN~2008ax, and SN~2011dh (M$_{O}\sim0.3-0.75$ \Msun, for  $\beta_{ratio}=1$ and $\beta_{6300,6364}=0.5$; \citetalias{Jerkstrand15}), we note that SN~2017ivv has an \ion{O}{I} mass in the SN~IIb range (column 7, Table~\ref{temp}). Unfortunately, fast-declining SNe~II do not have a mass estimation using this method. 
For $\beta_{6300,6364}=1$ (i.e. assuming optically thin emission), we can obtain the minimum mass of oxygen required to produce the observed \ion{[O}{I]}. With this assumption, the derived \ion{[O}{I]} masses are half of those presented in Table~\ref{temp}. We note that these oxygen masses are comparable to those obtained using the relation presented by \citet{Uomoto86}.

From Table~\ref{temp}, we also note a large range in the \ion{[O}{I]} masses. \citetalias{Jerkstrand15} discuss how time-dependent radiative transfer effects and/or deviations from LTE can lead to different results at different epochs. For IIb SNe, the optimum time to estimate the \ion{[O}{I]} mass is around 150 days; before this time, the lines can be affected by radiative transfer effects, and at later times, the \ion{[O}{I]} $\lambda5577$ line starts to deviate from LTE. For SN~2017ivv, at 158 days, we found a M(\ion{O}{I})=$0.66$ \Msun\ for $\beta_{ratio=1}$, M(\ion{O}{I})=$0.94$ \Msun\ for $\beta_{ratio=1.5}$ and M(\ion{O}{I})=$1.22$ \Msun\ for $\beta_{ratio=2}$, these values correspond to a 15 -- 19 \Msun\ progenitor \citep{Jerkstrand14}.

\subsection{[Ca II]/[O I] ratio}

\begin{figure}
\centering
\includegraphics[width=0.98\columnwidth]{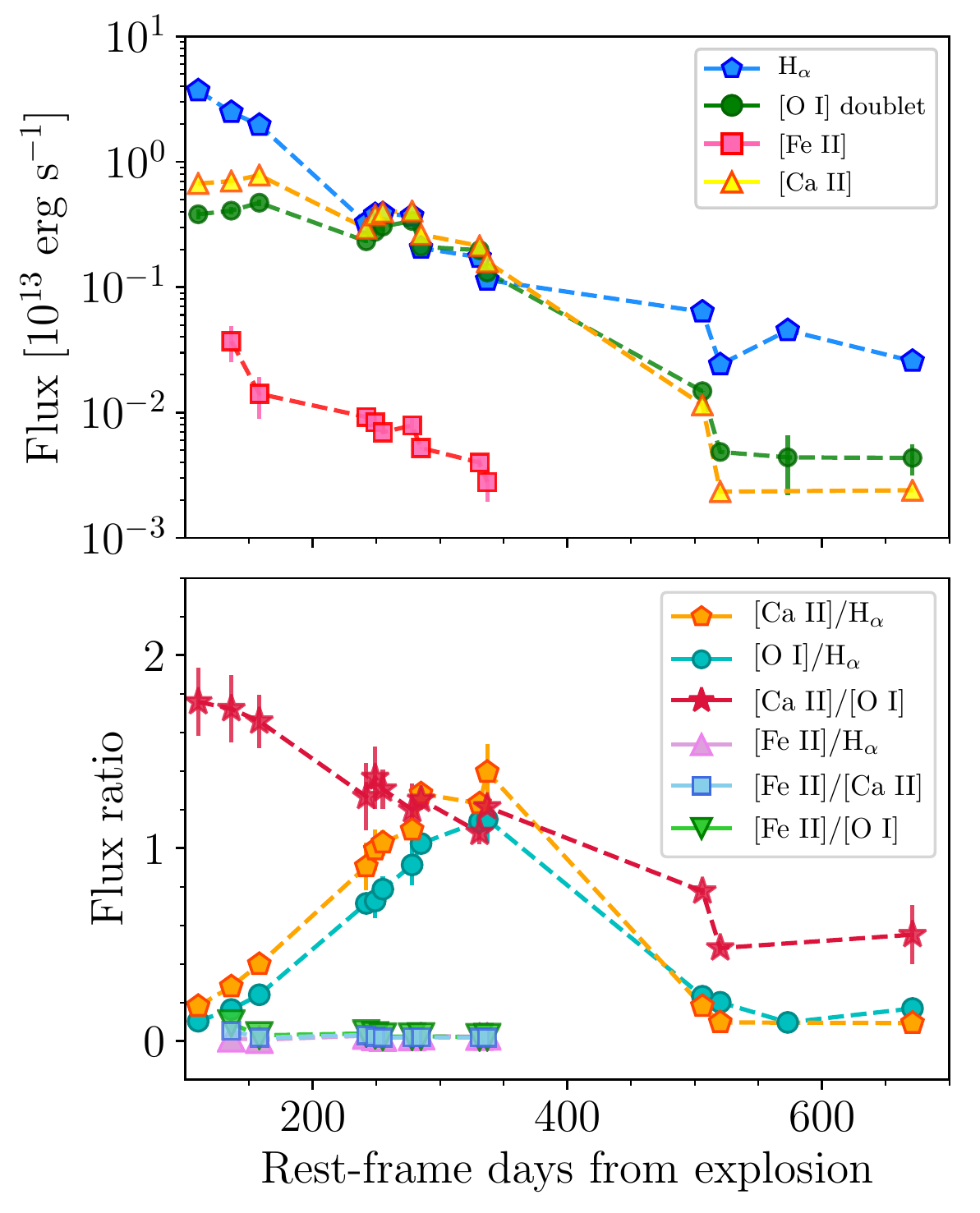}
\caption{\textbf{Top panel:} Flux evolution of H$\alpha$, \ion{[O}{I]}, \ion{[Ca}{II]}, and \ion{[Fe}{II]} in SN~2017ivv. \textbf{Bottom panel:} Flux ratio of the emission lines in SN~2017ivv.}
\label{flux}
\end{figure}

\citet{Fransson89} showed that the relative abundances of the different elements vary substantially with the mass of the helium core. Specifically, the \ion{[Ca}{II]}/\ion{[O}{I]} ratio can be used to estimate the ZAMS mass of the progenitor. Several authors \citep[e.g.][]{Fransson89, Elmhamdi04,Kuncarayakti15} have used this ratio as a diagnostic of the core mass. However, some issues emerge from these assumptions. \citet{Li93} demonstrated that in H-rich SNe most of the emission tends to come from primordial calcium in the hydrogen envelope. In SNe~IIb, the H envelopes are too small for such a contribution, and the Ca emission comes from the explosive oxygen burning zone. The mass of this zone depends on the explosion energy
\citep{Jerkstrand17}, and therefore the link between \ion{[Ca}{II]}/\ion{[O}{I]} ratio and the core mass is complex. 

\begin{figure}
\centering
\includegraphics[width=\columnwidth]{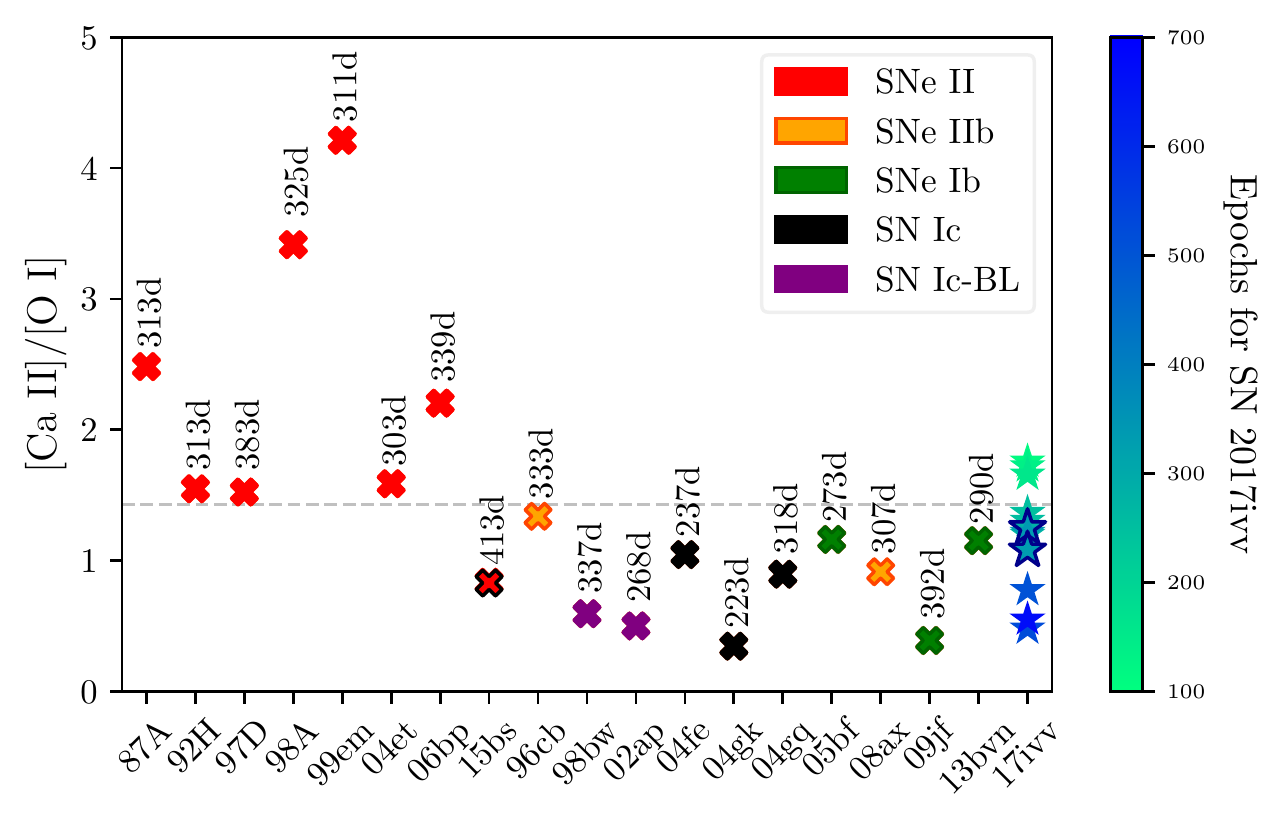}
\caption{\ion{[Ca}{II]}/\ion{[O}{I]} line ratio for several CC-SNe at similar ages during the nebular phase. The SN phases are computed with respect to the time of maximum light. Red crosses indicate SN~II, orange SN~IIb, green SN~Ib, black SN~Ic, and purple broad-lined Type-Ic. SN~2017ivv is shown with star symbols. SN~2017ivv is colour-coded respect to the epoch from the explosion. Stars with darker contours show
similar age than the comparison sample. The horizontal dashed line indicates the limit between SNe~II and stripped envelope events ($\sim1.43$). Figure adapted from \citet{Kuncarayakti15}.}
\label{caoi}
\end{figure}

Taking into account these caveats, we estimate the flux of these emission lines of SN~2017ivv and their ratios in Figure~\ref{flux}. From the top panel,  
one sees that all lines in SN~2017ivv evolve with time. The most extreme evolution is seen in the H$\alpha$ line; \ion{[Ca}{II]} and \ion{[O}{I]} evolve more slowly. Their ratio, shown in the bottom panel of Figure~\ref{flux}, decreases from $1.76\pm0.17$ at 110 days to $0.55\pm0.15$ at 671 days. Between 110 and 158 days, the ratio is larger than 1.5, which is consistent with the values presented for SNe~II by \citet{Kuncarayakti15}, while after 242 days the values are smaller and comparable to those of stripped envelope SNe \citep{Kuncarayakti15}. In Figure~\ref{caoi}, we compare the line ratio for SN~2017ivv with that of other CC-SNe. The ratio at 285 and 331 days is highlighted with darker contours showing similar age than the comparison sample. Although the ratio differs by $\sim0.2$, they are still consistent with the values of the stripped envelope SNe. The ratio of the slow-declining SN~II 2015bs also lies in this region, but at later epoch than other comparison objects.

The evolution of the \ion{[Ca}{II]}/\ion{[O}{I]} ratio in SN~2017ivv carries problems for any estimate of mass of the helium core: if we take the ratio at earlier phases (before 200 days), we are led to conclude that the SN has a low helium core mass (larger ratio, smaller masses). However, if we consider the ratio after 200-300 days, the result implies a more massive helium core. 
\citet{Valenti14} and \citet{Terreran19}, using a large sample of CC-SNe, found that this line ratio generally evolves with time, and \citet{Jerkstrand17} show that stripped-envelope models typically predict a declining ratio with time. Using their figure 13, the observed ratio of $\sim$1 at 300d most closely matches 3-4 \Msun~He core models, which would correspond to a $M_{ZAMS}\lesssim$15 \Msun\ progenitor (if its a IIb SN). However, as they discuss, the ratio is sensitive to various parameters (e.g. explosion energy, mixing, composition details).

\subsection{Blue-shifted lines}

\begin{figure}
\centering
\includegraphics[width=0.98\columnwidth]{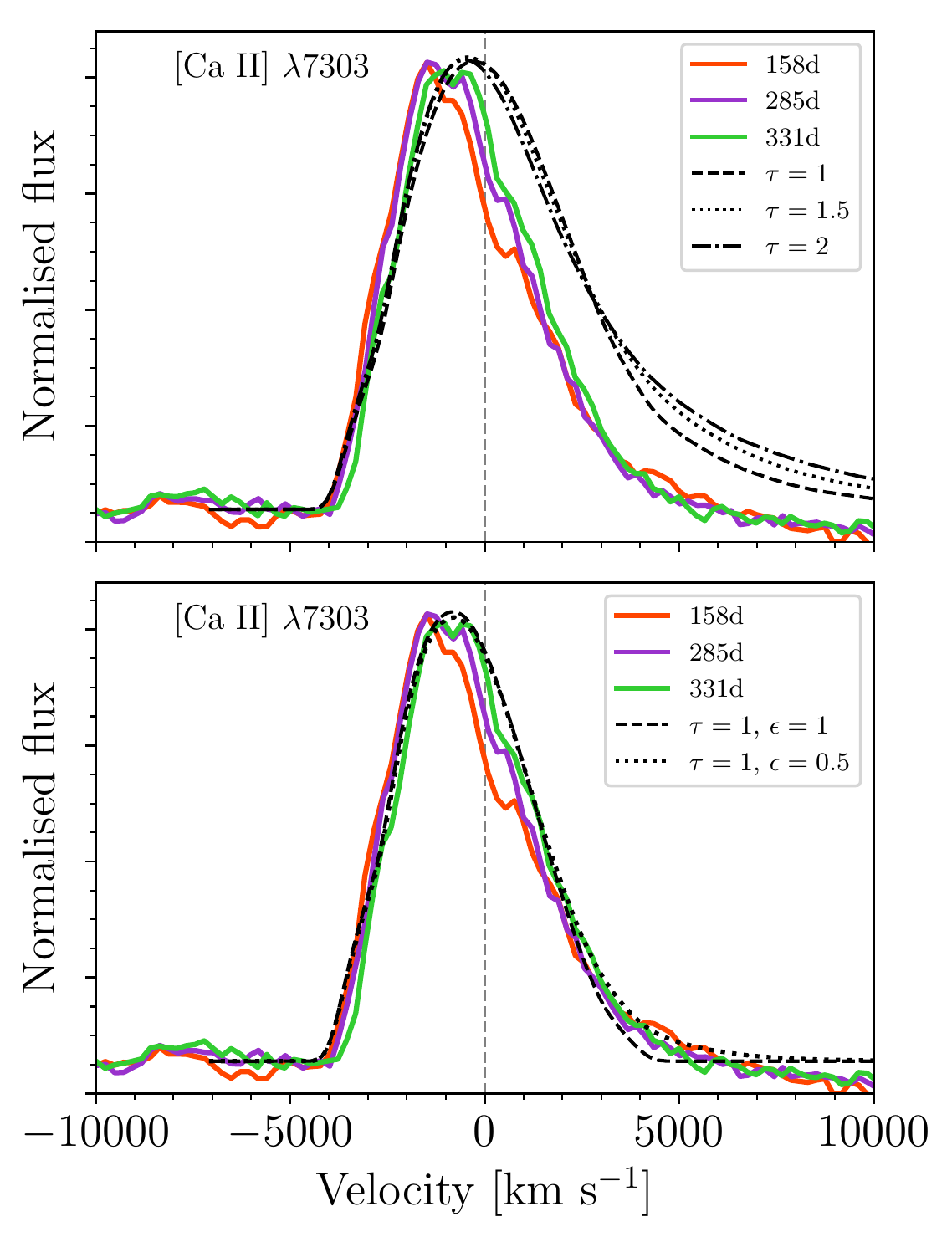}
\caption{\textbf{Top panel:} Observed \ion{[Ca}{II]} line at 158, 285 and 331 days compared to models of a uniform sphere with different degree of scattering optical depth. \textbf{Bottom panel:} Observed \ion{[Ca}{II]} line at 158, 285 and 331 days compared with a fully destructive ($\epsilon=1$) and partially destructive ($\epsilon=0.5$) opacity model. The zero velocity of \ion{[Ca}{II]} is at 7303 \AA. The models are normalised to match the peak flux.}
\label{modelca2}
\end{figure}

We investigate whether opacity effects can produce the observed asymmetric lines profiles in SN 2017ivv by modelling a uniform sphere with a velocity of 3500 \kms, with the optical depth $\tau$ and destruction probability $\epsilon$ as the two parameters.
Starting with models having pure scattering ($\epsilon=0$), no choice of optical depth produces the right combination of a blue-shifted peak and red wing for \ion{[Ca}{II]} (Figure~\ref{modelca2}, top panel).   Although scattering leads to a blue-shifting of the peak, it is not sufficient and the red-wing is overproduced. 

Moving on to explore models where (partial) destruction is allowed for, we find that 
for $\epsilon\sim 0.5$ (50\% destruction, 50\% scattering), the line profiles come into quite good agreement (Figure~\ref{modelca2}, bottom panel). In this model, some scattering is still needed to make the ``red tail'' of the lines, which extends up to about 6000 \kms, while absorption is needed to push the peak sufficiently to the blue (absorption shifts the peak more strongly than scattering).

This exercise demonstrated that the asymmetric line profiles can in principle be produced by radiative transfer effects, even in a simple toy model, and thus the necessity of an asymmetric ejecta cannot be directly established without more detailed modelling. The most plausible candidate for this opacity is line blocking which has been demonstrated to operate in stripped-envelope SNe up to $\sim$200d \citepalias{Jerkstrand15}. Depending on whether absorbed photons resonance scatter or fluoresce (true thermalization is unimportant), various epsilon values can be conceived of and a value of around 0.5 as indicated here is plausible. The main challenge for this mechanism is to sustain significant opacity for long enough. In the \citetalias{Jerkstrand15} models the optical depths became too small to significantly affect the lines profiles after about 200d. That also corresponds to the epoch around which line peak blue-shifts evolution ceased in many Ib/IIb SNe.

Current 3D SN explosion simulations \citep[e.g.,][]{Wongwathanarat13,Wongwathanarat15} predict asymmetries in line profiles with line centroid shifts up to about $\pm$500 \kms~for the IIP case \citep{Jerkstrand20}. For stripped envelope SNe, the scale of these shifts are likely yet larger due to overall higher velocities. Thus, it would not be surprising to see lines shifted by order $\pm1000$ \kms~in a stripped-envelope SN. SN 1993J and SN 2008ax settled on line peak shifts of about $-500$ \kms (blue-shifts), whereas SN 2011dh settled on almost symmetric or slightly reds-hifted lines \citep{Ergon14,Ergon15}. The settling of 2017ivv on 1000-2000 \kms~blue-shifts is thus quite remarkable, but it is perhaps not out of the ballpark of what current standard explosion models predict for a viewing angle close to the bulk momentum vector of the ejecta.

\subsection{SN progenitor}
\label{smodels}

\begin{figure*}
\centering
\includegraphics[width=8.5cm]{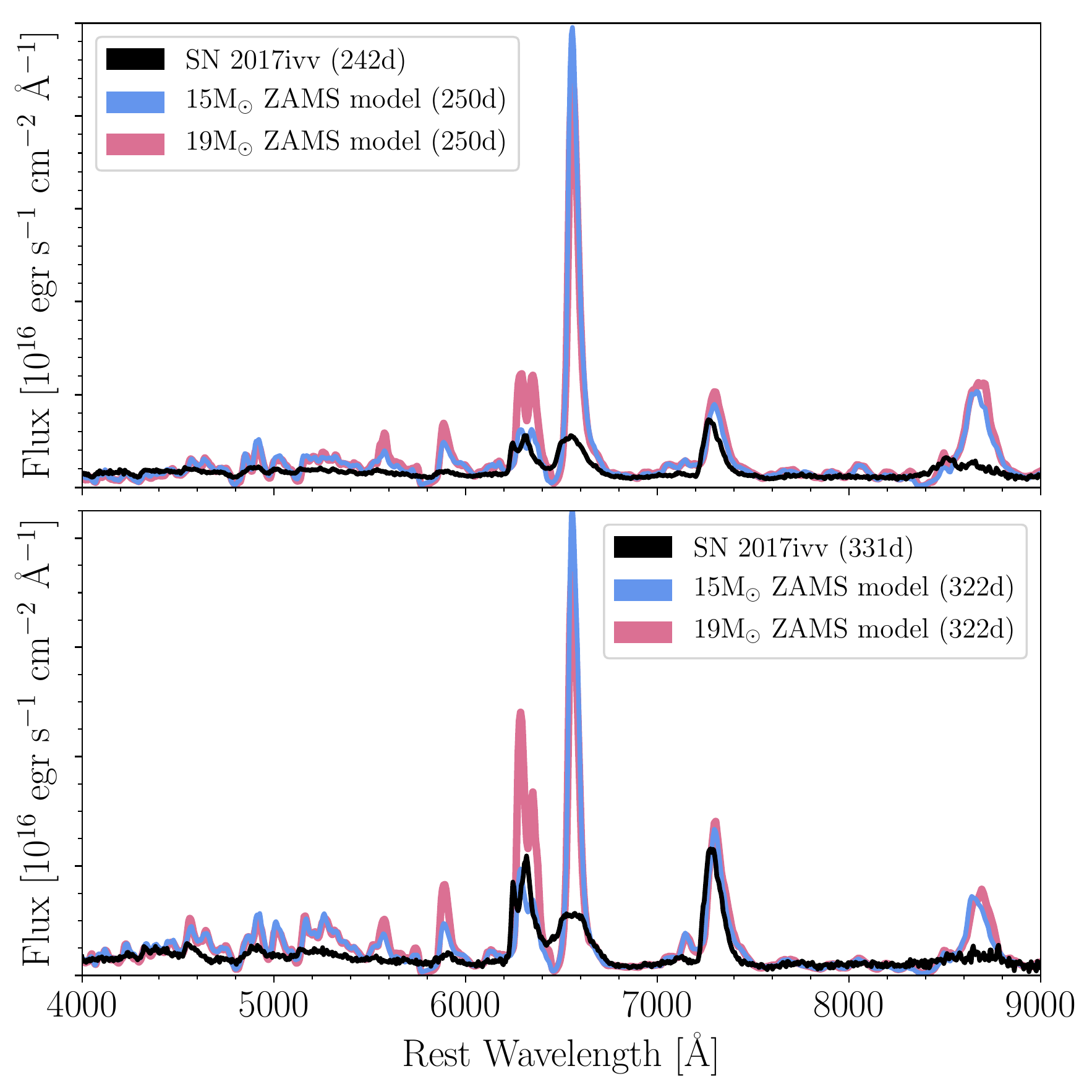}
\includegraphics[width=8.5cm]{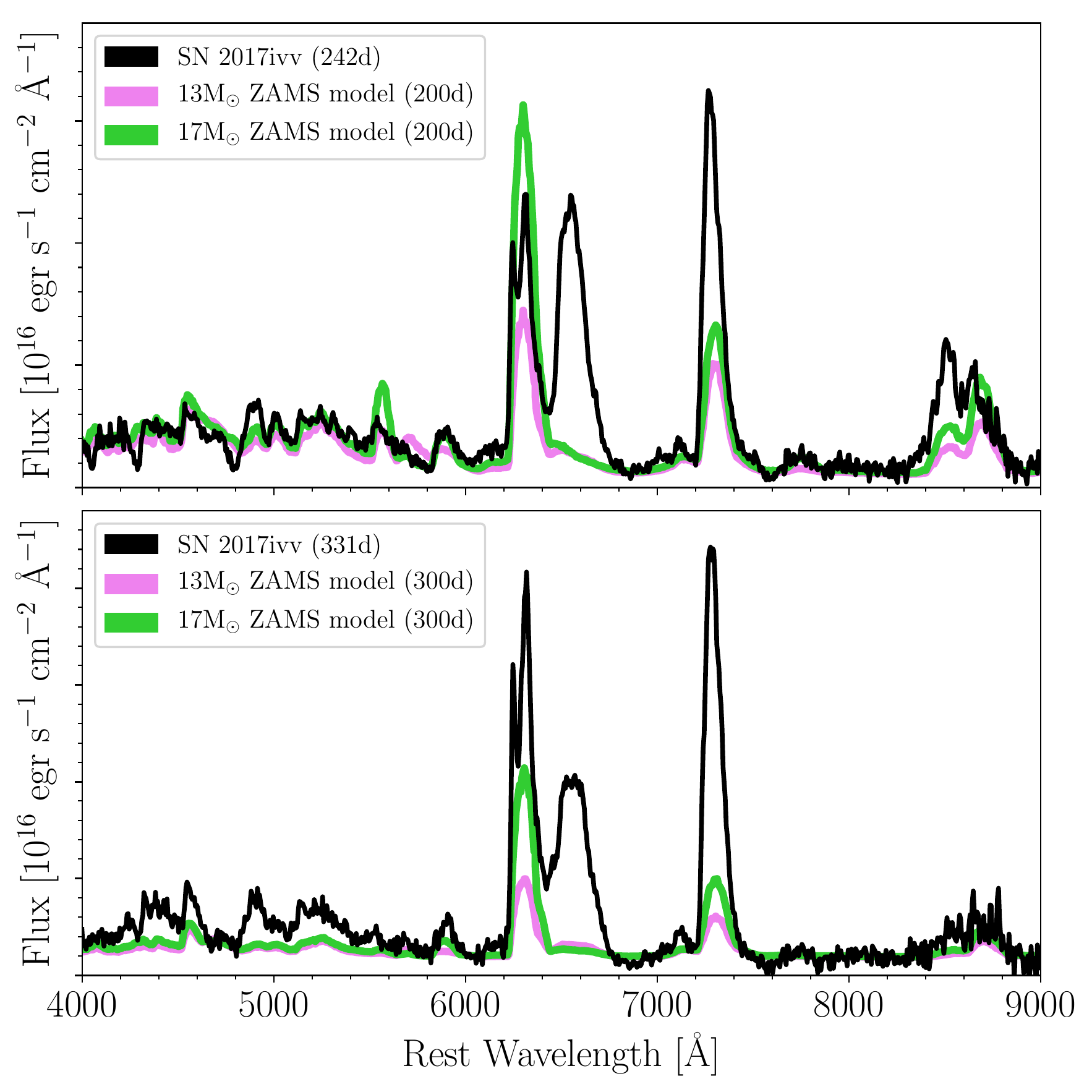}
\caption{Comparison of the nebular spectra of SN~2017ivv at 331 days with the nebular models. \textbf{Left panel:} Comparison with SN~II models of M$_{ZAMS}$ of 15 and 19 \Msun \citep{Jerkstrand14}. \textbf{Right panel:} Comparison with a SN~IIb model of M$_{ZAMS}=13$ (13G) and 17 \Msun. The models were scaled to same distance and same $^{56}$Ni mass as SN 2017ivv. }
\label{modelcomp}
\end{figure*}

After 500 days, the spectra and light curves of SN~2017ivv show significant signs of interaction between the SN ejecta and the CSM. These signs suggest that the SN progenitor must have lost a large amount of its hydrogen envelope before the explosion. Given that SN~2017ivv exploded in a low-metallicity environment, several considerations must be taken. If we assume a relation between metallicity and mass-loss \citep{Heger03}, at 
low-metallicity the mass-loss through winds should be less efficient. This may make the binary progenitor scenario more likely. On the other hand, if we consider the mass loss of single RSGs appears to be independent of metallicity \citep{vanLoon05,Goldman17,Chun18,Gutierrez18}, the metallicity-driven winds are still possible. However, a more massive progenitor ($>20$ \Msun) is maybe needed \citep[][and references therein]{Claeys11}.

We tried to fit the light curve using hydrodynamical simulations \citep{Limongi20} starting from "realistic" pre-SN models \citep{Limongi18} by changing the explosion energy. None of these simulations were able to reproduce the observations. The only way to fit the data was by artificially reducing the ejecta mass to values as low as $\sim0.8$ \Msun. This confirms that the binary scenario is the more likely.

To constrain the progenitor mass, we use the late-time spectral models of \citet{Jerkstrand14}, \citetalias{Jerkstrand15}.  We compare two spectra of SN~2017ivv, at 242 and 331 days with the 15 and 19 \Msun\ models for SNe~II \citep{Jerkstrand14}, and 13 and 17 \Msun\ models for SNe~IIb  \citepalias{Jerkstrand15}. As the strength of the \ion{[O}{I]} $\lambda\lambda6300,$ 6364 shows a correlation with the progenitor ZAMS mass \citep{Jerkstrand12,Jerkstrand14}, a match with the \ion{[O}{I]} lines will provide good constraints. To scale the models, we used the relation (2) from \citet{Bostroem19}
\begin{center}
$\frac{F_{obs}}{F_{mod}}=\frac{d^2_{mod}}{d^2_{obs}}\frac{M(\nifs)_{obs}}{M(\nifs)_{mod}}exp\left(\frac{t_{mod}-t_{obs}}{114.4}\right)$   
\end{center}

{\noindent} where F$_{obs}$ and F$_{mod}$ are the observed and models fluxes, d$^2_{obs}$ and d$^2_{mod}$ the observed and model distances, M(\nifs)$_{obs}$ and M(\nifs)$_{mod}$, the observed and model \nifs\ masses, and t$_{obs}$ and t$_{mod}$ are the phases of the observation and the model. The comparison is presented in Figure~\ref{modelcomp}. At 242 days, we find that the \ion{[O}{I]} strength matches with the 15 \Msun\ model for SNe~II, while at 331 days, the \ion{[O}{I]} strength falls between the 15 and 19 \Msun. However, the severe discrepancy with the H$\alpha$ fluxes makes it difficult to put much weight on these numbers.
In comparisons with SN~IIb models, the discrepancy goes the other way (SN 2017ivv likely has more H than these models) but is arguably less severe. We find that at 242 days, \ion{[O}{I]} strength falls between the 13 and 17 \Msun\ model, but at 331 days, the strength of \ion{[O}{I]} is larger than the 17 \Msun model. The IIb models have a too rapidly increasing gamma-ray escape compared to SN~2017ivv, as clarified by them being significantly too dim at 330 days. Thus the 240 days comparison (where the trapping is more in agreement) is more relevant than the 330d one. Given that the \ion{[O}{I]} $\lambda\lambda6300,$ 6364 doublet shows an unusual evolution in SN~2017ivv, the constraints on the progenitor mass must be analysed with caution. However, based on the spectral comparison, we estimate a progenitor between 15 -- 17 \Msun. 

\section{Summary and conclusions}
\label{conclusions}

We have presented the photometric and spectroscopic observations of the core collapse-supernova, SN~2017ivv. SN~2017ivv exploded in the faintest host galaxy (M$_r=-10.3$ mag) reported for a SN~II to date. 
Analysing the observed properties at early and late times, we found that SN~2017ivv shares properties with the fast-declining SN~II and SN~IIb. Based on the unusual evolution and the emission lines during the nebular phase, we suggest that SN~2017ivv has a lower hydrogen mass than a typical SN~II but more than a SN~IIb.

At early times, the light curve of SN~2017ivv shows a fast rise (6 -- 8 days) to the peak (${\rm M}^{\rm max}_{g}= -17.84$ mag), followed by a very rapid decline ($7.94\pm0.48$ mag per 100 days in the $V-$band). The spectra show broad P-Cygni profiles, with H$\alpha$ exhibiting an asymmetric emission component and a weak absorption feature.
At late times, based on the  spectroscopic behaviour, we could split the evolution into three phases (1) H$\alpha$ strong phase ($<200$ days); (2) H$\alpha$ weak phase (between 200 and 350 days); and (3) H$\alpha$ broad phase ($>500$ days).  
In phase one, SN~2017ivv shows the typical spectra of a SN~II: a dominant H$\alpha$ emission profile, and clear detections of the \ion{[O}{I]} and  \ion{[Ca}{II]} lines. In phase two, H$\alpha$ develops a boxy emission profile, while \ion{[O}{I]} and \ion{[Ca}{II]} become the strongest lines in the spectra. The \ion{[O}{I]} $\lambda\lambda6300,$ 6364 displays an unusual profile with the red peak stronger than the blue one. These  properties, observed before in some stripped-envelope SNe, can be associated with aspherical explosions. Finally, in phase three, the spectra display weak \ion{[O}{I]} and \ion{[Ca}{II]} lines, combined with a broad and strong H$\alpha$ emission. These characteristics are likely explained by CSM interaction, which is also supported by the flattening in the light curve at late-times.

Taking into account all these properties, we concluded that SN~2017ivv arises from an asymmetric explosion and that its progenitor had a significant mass-loss prior to the explosion. As it explodes in a low-metallicity environment, a binary progenitor system is the most likely scenario. From the nebular analysis, we infer a progenitor mass of 15 -- 17 \Msun.

\section*{Acknowledgements}

We thank the anonymous referee for the comments and suggestions that have helped to improve the paper.

We are grateful to Marco Limongi and Giacomo Terreran for useful discussion.
We  thank  Richard S. Post its contribution with data from the Post Observatory.

C.P.G. and M.S. acknowledge support from EU/FP7-ERC grant No. [615929].
A.J. acknowledges funding by ERC Starting Grant 803189 and Swedish National
Research Council grant 2018-03799.
L.G. was funded by the European Union's Horizon 2020 research and innovation programme under the Marie Sk\l{}odowska-Curie grant agreement No. 839090.
M.F. is supported by a Royal Society - Science Foundation Ireland University Research Fellowship.
K.M. acknowledges support from ERC Starting Grant grant no. 758638.
S.J.S. acknowledges funding from STFC Grant Ref: ST/P000312/1. 
T.M.B. was funded by the CONICYT PFCHA / DOCTORADOBECAS CHILE/2017-72180113. 
T.W.C. acknowledges EU Funding under Marie Sk\l{}odowska-Curie grant agreement No 842471
M.G. is supported by the Polish NCN MAESTRO grant 2014/14/A/ST9/00121.
D.A.H. and G.H. were supported by NSF grant AST-1313484.
M.N. is supported by a Royal Astronomical Society Research Fellowship.
Support for G.P. and J.L.P. is provided by the Ministry of Economy, Development, and Tourism's Millennium Science Initiative through grant IC120009, awarded to The Millennium Institute of Astrophysics, MAS.
J.L.P. is also supported by FONDECYT through the grant 1191038. 
Research by K.A.B and S.V. is supported by NSF grant AST--1813176. 
C.S.K. is supported by NSF grants AST-1515927 and AST-181440. C.S.K. and B.J.S. are supported by NSF grant AST-1907570.  B.J.S. is also supported by NSF grants AST-1920392 and AST-1911074. J.B., D.H., And D.A.H. are supported by NSF NSF AST-1911225.

This work used data collected at the European Organisation for Astronomical Research in the Southern Hemisphere, Chile, under program IDs: 0104.D-0503(A), 0103.D-0393(A), 0103.D-0440(A), 0102.D-0356(A), 0102.A-9099(A), 0101.A-9099(A), and as part of PESSTO,
(the Public ESO Spectroscopic Survey for Transient Objects Survey) ESO program 1103.D-0328 and 199.D-0143.

This work makes use of data from Las Cumbres Observatory telescope network. 

Part of the funding for GROND (both hardware as well as personnel) was generously granted from the Leibniz-Prize to Prof. G. Hasinger (DFG grant HA 1850/28-1). 

This research uses data obtained through the Telescope Access Program (TAP), which has been funded by the National Astronomical Observatories of China, the Chinese Academy of Sciences, and the Special Fund for Astronomy from the Ministry of Finance.

Partially based on observations collected at Copernico 1.82m and Schmidt 67/92 telescopes (Asiago, Italy) of the INAF - Osservatorio Astronomico di Padova.

The Liverpool Telescope is operated on the island of La Palma by Liverpool John Moores University in the Spanish Observatorio del Roque de los Muchachos of the Instituto de Astrofisica de Canarias with financial support from the UK Science and Technology Facilities Council.

Some of the data presented herein were obtained at the W. M. Keck Observatory, which is operated as a scientific partnership among the California Institute of Technology, the University of California and the National Aeronautics and Space Administration. The Observatory was made possible by the generous financial support of the W. M. Keck Foundation.

We thank the Las Cumbres Observatory and its staff for its continuing support of the ASAS-SN project. ASAS-SN is supported by the Gordon and Betty Moore Foundation through grant GBMF5490 to the Ohio State University, and NSF grants AST-1515927 and AST-1908570. Development of ASAS-SN has been supported by NSF grant AST-0908816, the Mt. Cuba Astronomical Foundation, the Center for Cosmology and AstroParticle Physics at the Ohio State University,  the Chinese Academy of Sciences South America Center for Astronomy (CAS- SACA), the Villum Foundation, and George Skestos. 

This work has made use of data from the Asteroid Terrestrial-impact Last Alert System (ATLAS) project. ATLAS is primarily funded to search for near earth asteroids through NASA grants NN12AR55G, 80NSSC18K0284, and 80NSSC18K1575; by products of the NEO search include images and catalogs from the survey area. The ATLAS science products have been made possible through the contributions of the University of Hawaii Institute for Astronomy, the Queen's University Belfast, the Space Telescope Science Institute, and the South African Astronomical Observatory.

This work has been partially supported by the Spanish grant PGC2018-095317-B-C21 within the European Funds for Regional Development (FEDER).

This research has made use of the NASA/IPAC Extragalactic Database (NED) which is operated by the Jet Propulsion Laboratory, California Institute of Technology, under contract with the National Aeronautics. We acknowledge the usage of the HyperLeda database (\url{http://leda.univ-lyon1.fr})

\textit{Facilities:} NTT (EFOSC2); VLT (MUSE, FORS2); Las Cumbres Observatory; ATLAS; Keck: I (LRIS); 1.82 m Copernico (AFOSC); 67/91 Schmit Telescope; 1.22 Galileo (BC); ASAS-SN; Liverpool Telescope (SPRAT); Tillinghast (FAST); 2.2-m MPG telescope (GROND); Post Observatory SRO; 

\textit{Software:} Python from \url{https://www.python.org/}, IRAF, \texttt{EsoReflex} pipeline \citep{Freudling13}; \sc{LPIPE} \citep{Perley19}; GROND pipeline \citep{Kruhler08}, \textsc{snid} \citep{Blondin07}.




\bibliographystyle{mnras}
\bibliography{Bibliography}


\appendix

\section{Tables}
\label{ap1}

\renewcommand{\thetable}{A\arabic{table}}
\setcounter{table}{0}
\begin{table}
\centering
\begin{threeparttable}
\caption{ATLAS AB photometry.}
\label{photoatlas}
\begin{tabular}[t]{ccccccc}
\hline
\hline
UT date    &	MJD	    & Phase           &  \textit{orange} filter  \\
           &            & (days)$^{\ast}$ &  (mag)     \\
\hline                                
\hline                                
20171205   & 58092.22   &    0.60    & $15.26\pm0.01$ \\
20171209   & 58096.22   &    4.57    & $14.35\pm0.01$ \\
20171211   & 58098.23   &    6.57    & $14.29\pm0.01$ \\
20171213   & 58100.22   &    8.55    & $14.32\pm0.01$ \\
20180329   & 58206.63   &    114.37  & $16.50\pm0.03$ \\
20180520   & 58258.50   &    165.95  & $18.05\pm0.08$ \\
20180526   & 58264.60   &    172.02  & $17.39\pm0.04$ \\
20180531   & 58269.53   &    176.92  & $17.39\pm0.08$ \\
20180611   & 58280.61   &    187.94  & $17.39\pm0.11$ \\
20180615   & 58284.60   &    191.91  & $17.50\pm0.05$ \\
20180617   & 58286.60   &    193.89  & $18.80\pm0.11$ \\
20180621   & 58290.58   &    197.85  & $17.78\pm0.04$ \\
20180623   & 58292.55   &    199.81  & $17.79\pm0.05$ \\
20180625   & 58294.55   &    201.80  & $17.83\pm0.06$ \\
20180626   & 58295.45   &    202.69  & $17.76\pm0.11$ \\
20180627   & 58296.51   &    203.75  & $17.76\pm0.18$ \\
20180707   & 58306.57   &    213.75  & $17.99\pm0.07$ \\
20180717   & 58316.51   &    223.64  & $18.08\pm0.07$ \\
20180721   & 58320.51   &    227.62  & $18.13\pm0.09$ \\
20180723   & 58322.51   &    229.60  & $18.26\pm0.07$ \\
20180725   & 58324.47   &    231.55  & $18.18\pm0.14$ \\
20180806   & 58336.49   &    243.51  & $18.36\pm0.09$ \\
20180810   & 58340.47   &    247.46  & $18.36\pm0.12$ \\
20180814   & 58344.46   &    251.43  & $18.45\pm0.11$ \\
20180818   & 58348.44   &    255.39  & $18.52\pm0.13$ \\
20180830   & 58360.43   &    267.31  & $18.78\pm0.16$ \\
20180901   & 58362.42   &    269.29  & $18.60\pm0.12$ \\
20180903   & 58364.41   &    271.27  & $18.67\pm0.13$ \\
20180909   & 58370.40   &    277.23  & $19.59\pm0.13$ \\
20180911   & 58372.39   &    279.21  & $18.82\pm0.14$ \\
20180915   & 58376.37   &    283.16  & $19.02\pm0.15$ \\
20180917   & 58378.38   &    285.16  & $18.85\pm0.13$ \\
20180929   & 58390.35   &    297.07  & $19.04\pm0.16$ \\
20181001   & 58392.37   &    299.08  & $18.98\pm0.18$ \\
20181003   & 58394.35   &    301.04  & $18.58\pm0.13$ \\
20181015   & 58406.32   &    312.95  & $19.23\pm0.17$ \\
\hline
\end{tabular}
\begin{list}{}{}
\item \textbf{Notes:} 
\item $^{\ast}$ Rest-frame phase in days from explosion. 
\end{list}
\end{threeparttable}
\end{table}

\renewcommand{\thetable}{A\arabic{table}}
\setcounter{table}{1}
\begin{table*}
\centering
\caption{Optical photometry from Las Cumbres Observatory.}
\label{photolco}
\begin{tabular}[t]{cccccccccc}
\hline
\hline
UT date    &	MJD	    & Phase           &       $B$	    &       $V$	     &       $g$	   &	$r$	        &	$i$	         \\
           &            & (days)$^{\ast}$ &     (mag)      &     (mag)      &     (mag)       &     (mag)      &     (mag)      \\
\hline
\hline
20171214   &  58101.20	&  9.53          & $14.66\pm0.06$ & $14.42\pm0.03$ & \nodata         & $14.41\pm0.04$ & $14.41\pm0.06$ \\
20180316   &  58193.40	&  101.21        & $18.56\pm0.03$ & $17.43\pm0.02$ & \nodata         & $16.44\pm0.01$ & $16.70\pm0.02$ \\
20180323   &  58200.80	&  108.57        & $18.55\pm0.04$ & $17.46\pm0.02$ & $17.88\pm0.015$ & $16.51\pm0.01$ & $16.77\pm0.01$ \\
20180401   &  58209.40	&  117.12        & $18.73\pm0.06$ & $17.57\pm0.02$ & $17.98\pm0.022$ & $16.61\pm0.01$ & $16.88\pm0.01$ \\
20180411   &  58219.80	&  127.47        & $18.87\pm0.04$ & $17.71\pm0.02$ & $18.16\pm0.018$ & $16.75\pm0.01$ & $17.01\pm0.02$ \\
20180420   &  58228.10  &  135.72        & \nodata        & $17.91\pm0.03$ & \nodata         & $16.87\pm0.01$ & $17.11\pm0.02$ \\
20180427   &  58235.70	&  143.28        & $19.12\pm0.12$ & $17.93\pm0.04$ & $18.28\pm0.040$ & $16.91\pm0.02$ & $17.17\pm0.02$ \\
20180505   &  58243.40	&  150.93        & $19.12\pm0.06$ & $18.10\pm0.03$ & $18.45\pm0.029$ & $17.06\pm0.01$ & $17.27\pm0.02$ \\
20180513   &  58251.30	&  158.79        & $19.02\pm0.06$ & $18.16\pm0.03$ & $18.49\pm0.027$ & $17.20\pm0.01$ & $17.44\pm0.02$ \\
20180523   &  58261.40	&  168.83        & $19.29\pm0.02$ & $18.37\pm0.01$ & $18.65\pm0.010$ & $17.28\pm0.01$ & $17.47\pm0.01$ \\
20180602   &  58271.30	&  178.68        & $19.50\pm0.10$ & $18.40\pm0.04$ & $18.78\pm0.038$ & $17.38\pm0.02$ & $17.63\pm0.02$ \\
20180612   &  58281.10	&  188.42        & $19.49\pm0.04$ & $18.65\pm0.03$ & $18.88\pm0.020$ & $17.53\pm0.01$ & $17.72\pm0.02$ \\
20180621   &  58290.80	&  198.07        & $19.55\pm0.02$ & $18.75\pm0.01$ & $19.01\pm0.011$ & $17.70\pm0.01$ & $17.81\pm0.01$ \\
20180629   &  58298.10	&  205.33        & $19.19\pm0.13$ & \nodata        & \nodata         & \nodata        & \nodata        \\
20180629   &  58298.20	&  205.43        & $19.03\pm0.12$ & $18.72\pm0.12$ & $19.37\pm0.146$ & $17.84\pm0.05$ & $17.84\pm0.06$ \\
20180707   &  58306.30	&  213.48        & $19.73\pm0.05$ & $18.88\pm0.03$ & \nodata         & \nodata        & \nodata        \\
20180707   &  58306.40  &  213.58        & \nodata        & $18.95\pm0.04$ & $19.19\pm0.027$ & $17.90\pm0.02$ & $18.02\pm0.02$ \\
20180715   &  58314.70	&  221.84        & $19.82\pm0.02$ & $19.06\pm0.02$ & $19.27\pm0.012$ & $18.04\pm0.01$ & $18.08\pm0.01$ \\
20180723   &  58322.30	&  229.40        & $19.96\pm0.09$ & $19.15\pm0.05$ & $19.43\pm0.061$ & $18.16\pm0.03$ & $18.20\pm0.04$ \\
20180724   &  58323.70	&  230.79        & $19.92\pm0.07$ & $19.19\pm0.04$ & $19.50\pm0.037$ & $18.18\pm0.01$ & $18.22\pm0.02$ \\
20180811   &  58341.60	&  248.59        & $19.96\pm0.05$ & $19.30\pm0.04$ & $19.43\pm0.024$ & $18.39\pm0.02$ & $18.41\pm0.02$ \\
20180827   &  58357.20  &  264.10        & \nodata        & \nodata        & \nodata         & $18.55\pm0.10$ & \nodata        \\
20180912   &  58373.10	&  279.91        & $20.35\pm0.05$ & $19.82\pm0.04$ & $19.99\pm0.026$ & $18.85\pm0.02$ & $18.80\pm0.02$ \\
20180929   &  58390.80	&  297.51        & $20.62\pm0.08$ & $19.91\pm0.07$ & $20.12\pm0.044$ & $19.02\pm0.03$ & $19.09\pm0.05$ \\
20181023   &  58414.40	&  320.98        & $21.04\pm0.19$ & $20.37\pm0.11$ & $20.38\pm0.090$ & $19.30\pm0.04$ & $19.37\pm0.05$ \\
\hline
\end{tabular}
\begin{list}{}{}
\item \textbf{Notes:} \\
$^{\ast}$ Rest-frame phase in days from explosion.
\end{list}
\end{table*}

\renewcommand{\thetable}{A\arabic{table}}
\setcounter{table}{2}
\begin{table*}
\centering
\scriptsize
\caption{Optical photometry from Asiago, Mount Ekar.}
\label{photoasiago}
\begin{tabular}[t]{cccccccccccc}
\hline
\hline
UT date    &	MJD	    & Phase           &       $B$	    &       $V$	     &       $u$	   &       $g$	    &	   $r$	        &	$i$	            &       $z$	 \\
           &            & (days)$^{\ast}$ &     (mag)      &     (mag)      &     (mag)       &     (mag)      &     (mag)         &     (mag)         &     (mag)   \\
\hline
\hline
20171216   & 58103.72 	&   12.03         & $14.68\pm0.02$ & $14.65\pm0.031$ & $14.44\pm0.01$ & $14.51\pm0.02$ & $14.424\pm0.018$  &  $14.516\pm0.016$ & $14.57\pm0.02$ \\
20171217   & 58104.69 	&   13.00         & $14.67\pm0.01$ & $14.64\pm0.009$ & $15.09\pm0.01$ & $14.65\pm0.02$ & $14.543\pm0.017$  &  $14.569\pm0.034$ & \nodata        \\ 
20171219   & 58106.71 	&   15.01         & $15.00\pm0.03$ & $14.74\pm0.044$ & \nodata 	   & $14.67\pm0.03$ & $14.745\pm0.071$  &  $14.702\pm0.029$ & $14.73\pm0.02$ \\
20171220   & 58107.71 	&   16.00         & $15.06\pm0.01$ & $14.87\pm0.013$ & $14.72\pm0.01$ & $14.89\pm0.01$ & $14.720\pm0.008$  &  $14.774\pm0.010$ & $14.77\pm0.01$ \\ 
20171222   & 58109.71 	&   17.99         & \nodata        & \nodata         & \nodata        & $14.97\pm0.05$ & $14.848\pm0.069$  &  $14.664\pm0.078$ & \nodata        \\ 
20171224   & 58111.71 	&   19.98         & $15.42\pm0.02$ & $15.22\pm0.026$ & $15.39\pm0.02$ & $15.20\pm0.02$ & $15.023\pm0.026$  &  $15.128\pm0.024$ & $15.10\pm0.02$ \\        
\hline
\end{tabular}
\begin{list}{}{}
\item \textbf{Notes:} \\
$^{\ast}$ Rest-frame phase in days from explosion.
\end{list}
\end{table*}

\renewcommand{\thetable}{A\arabic{table}}
\setcounter{table}{3}
\begin{table*}
\centering
\caption{Optical photometry from All-Sky Automated Survey for Supernovae.}
\label{photoasas}
\begin{tabular}[t]{cccccccccc}
\hline
\hline
UT date  &	MJD	    & Phase           &       $V$	   &       $g$	     \\
         &          & (days)$^{\ast}$ &     (mag)      &     (mag)       \\
\hline                                                                  
\hline                                
20171204 & 58091.03 &  \nodata        & \nodata        & $>17.14$        \\
20171212 & 58099.04 &  7.38           & \nodata        & $14.41\pm0.02$  \\    
20171213 & 58100.04 &  8.37           & \nodata        & $14.37\pm0.03$  \\
20171214 & 58101.04 &  9.37           & \nodata        & $14.63\pm0.03$  \\    
20180403 & 58211.47 &  119.18         & \nodata        & $17.45\pm0.29$  \\    
20180415 & 58223.36 &  131.01         & $17.47\pm0.24$ & \nodata         \\    
20180418 & 58226.46 &  134.09         & \nodata        & $18.49\pm0.33$  \\
20180514 & 58252.54 &  160.02         & $17.80\pm0.19$ & \nodata         \\    
20180516 & 58254.57 &  162.04         & $17.91\pm0.25$ & $18.64\pm0.31$  \\ 
20180519 & 58257.58 &  165.04         & $18.15\pm0.34$ & \nodata         \\    
20180520 & 58258.54 &  165.99         & $17.98\pm0.25$ & \nodata         \\    
20180521 & 58259.57 &  167.01         & $17.88\pm0.22$ & \nodata         \\    
20180715 & 58314.24 &  221.38         & $18.33\pm0.36$ & \nodata         \\ 
20180620 & 58289.29 &  196.57         & \nodata        & $18.96\pm0.33$  \\    
20180717 & 58316.19 &  223.32         & \nodata        & $18.70\pm0.36$  \\
\hline          
\end{tabular}
\begin{list}{}{}
\item \textbf{Notes:} 
\item $^{\ast}$ Rest-frame phase in days from explosion.
\end{list}
\end{table*}

\renewcommand{\thetable}{A\arabic{table}}
\setcounter{table}{4}
\begin{table*}
\centering
\caption{Optical photometry from the Post Observatory SRO.}
\label{photopost}
\begin{tabular}[t]{cccccccccc}
\hline
\hline
UT date  &	MJD	    & Phase           &       $B$	    &       $V$	    &	  $r$          & 	$i$	       \\
         &          & (days)$^{\ast}$ &     (mag)      &     (mag)      &     (mag)      &     (mag)      \\
\hline                                                                                      
\hline   
20171217 & 58104.09 &  12.40          & $14.87\pm0.05$ & $14.56\pm0.04$ & $14.53\pm0.04$ & $14.46\pm0.07$  \\
20171218 & 58105.13 &  13.43          & $15.03\pm0.07$ & $14.64\pm0.11$ & \nodata        & \nodata         \\    
20171220 & 58107.08 &  15.37          & $15.18\pm0.07$ & $14.83\pm0.05$ & $14.68\pm0.05$ & $14.70\pm0.11$  \\
20171221 & 58108.08 &  16.37          & $15.23\pm0.08$ & $14.89\pm0.07$ & $14.80\pm0.05$ & $14.72\pm0.11$  \\    
20171223 & 58110.08 &  18.36          & $15.22\pm0.11$ & $14.91\pm0.09$ & \nodata        & $14.79\pm0.11$  \\
\hline          
\end{tabular}
\begin{list}{}{}
\item \textbf{Notes:} 
\item $^{\ast}$ Rest-frame phase in days from explosion.
\end{list}
\end{table*}

\renewcommand{\thetable}{A\arabic{table}}
\setcounter{table}{5}
\begin{table*}
\centering
\caption{Optical and near-infrared photometry from  the  Gamma-Ray  Burst  Optical/Near-Infrared  Detector (GROND).}
\label{photogrond}
\begin{tabular}[t]{cccccccccc}
\hline
\hline
UT date  &	MJD	    & Phase           &       $r$	   &       $J$      &        $H$     &  $K$    \\
         &          & (days)$^{\ast}$ &     (mag)      &      (mag)     &       (mag)    & (mag)   \\
\hline                                                                  
\hline             
20180903 & 58363.06 &  271.44         & \nodata        & $18.61\pm0.06$ & $18.75\pm0.15$ & $>18.43$ \\ 
20180905 & 58366.14 &  274.52         & \nodata        & $18.69\pm0.06$ & $18.34\pm0.11$ & $>17.87$ \\
20180907 & 58368.06 &  276.44         & \nodata        & $18.54\pm0.05$ & $18.60\pm0.14$ & $>17.80$ \\
20190410 & 58583.35 &  488.99         & $21.01\pm0.11$ &  $>20.10$      &    $>19.35$    & \nodata  \\
20190920 & 58746.06 &  650.76         & $21.41\pm0.17$ &  $>20.30$      &    $>19.93$    & \nodata  \\
\hline          
\end{tabular}
\begin{list}{}{}
\item \textbf{Notes:} 
\item $^{\ast}$ Rest-frame phase in days from explosion.
\end{list}
\end{table*}

\renewcommand{\thetable}{A\arabic{table}}
\setcounter{table}{6}
\begin{table*}
\centering
\caption{Optical photometry from the Multi Unit Spectroscopic Explorer (MUSE).}
\label{photomuse}
\begin{tabular}[t]{cccccccccc}
\hline
\hline
UT date  &	MJD	    & Phase           &       $V$	   &       $r$	     &       $i$	  \\
         &          & (days)$^{\ast}$ &     (mag)      &     (mag)       &     (mag)      \\
\hline                                                                                   
\hline                                
20171204 & 58614.34  &  519.91        & $21.89\pm0.31$ & $21.15\pm0.13$  & $21.67\pm0.14$ \\
20171212 & 58766.00  &  670.62        & $22.11\pm0.28$ & $21.72\pm0.21$  & $22.13\pm0.09$  \\
\hline          
\end{tabular}
\begin{list}{}{}
\item \textbf{Notes:} 
\item $^{\ast}$ Rest-frame phase in days from explosion.
\end{list}
\end{table*}

\renewcommand{\thetable}{A\arabic{table}}
\setcounter{table}{7}
\begin{table*}
\centering
\caption{Spectroscopic observations of SN~2017ivv}
\label{tspectra}
\begin{tabular}[t]{cccccccccc}
\hline
\hline
UT date   &	MJD   & Phase       	&  Range  &  Telescope  & Grism/Grating & Resolution \\
          &           & [days]$^{\ast}$ &  [\AA]  & +Instrument &               &   [\AA]    \\
\hline
\hline
20171215  & 58102.07  &  10.39   & 3470-7400 &  Tillinghast+FAST &  300 gpm           &  6    \\  
20171215  & 58102.73  &  11.05   & 3400-8200 &  Ekar+AFOSC   	 &  gm4               &  19   \\  
20171216  & 58103.07  &  11.39   & 3470-7400 &  Tillinghast+FAST &  300 gpm           &  6     \\  
20171216  & 58103.70  &  12.01   & 3500-7300 &  Ekar+AFOSC       &  VPH7              &  11   \\  
20171216  & 58103.71  &  12.02   & 3400-7800 &  Pennar+BC        &  300tr/mm          &   10    \\
20171217  & 58104.72  &  13.03   & 3400-7800 &  Pennar+BC        &  300tr/mm          & 10      \\ 
20171219  & 58106.71  &  15.01   & 3400-8200 &  Ekar+AFOSC       &  gm4               &  19   \\
20171220  & 58107.70  &  15.99   & 3400-8200 &  Ekar+AFOSC       &  gm4               &  19   \\ 
20171221  & 58108.80  &  17.08   & 4000-8000 &  LT+SPRAT         &  VPH               &  18   \\  
20171222  & 58109.69  &  17.97   & 3400-8200 &  Ekar+AFOSC       &  gm4               &  19   \\  
20171224  & 58111.69  &  19.96   & 3400-8200 &  Ekar+AFOSC       &  gm4               &  19   \\  
20180325  & 58202.39  &  110.15  & 3700-9300 &  NTT+EFOSC2       &  Gr13              &  21   \\      
20180420  & 58228.35  &  135.97  & 3700-9300 &  NTT+EFOSC2       &  Gr13              &  21   \\
20180512  & 58250.27  &  157.77  & 3700-9300 &  NTT+EFOSC2       &  Gr13              &  21   \\   
20180805  & 58335.18  &  242.20  & 3700-9300 &  NTT+EFOSC2       &  Gr13              &  21   \\
20180812  & 58342.46  &  249.44  & 3100-10200&  Keck+LRIS        & 600/4000, 400/8500 &  6    \\
20180818  & 58348.19  &  255.14  & 3700-9300 &  NTT+EFOSC2       &  Gr13              &  21   \\
20180910  & 58371.17  &  277.99  & 3700-9300 &  NTT+EFOSC2       &  Gr13              &  21   \\
20180917  & 58378.11  &  284.89  & 3700-9300 &  NTT+EFOSC2       &  Gr13              &  21   \\
20181102  & 58424.01  &  330.54  & 3700-9300 &  NTT+EFOSC2       &  Gr13              &  21   \\
20181108  & 58430.02  &  336.52  & 4600-10000&  VLT+FORS2        &  GRIS$\_$150I      &  28   \\  
20190427  & 58600.35  &  505.90  & 3700-9300 &  NTT+EFOSC2       &  Gr13              &  21   \\
20190511  & 58614.34  &  519.81  & 4750-9350 &  VLT+MUSE         &  WFM$^{\star}$     &  3    \\
20190704  & 58668.28  &  573.45  & 4600-8850 &  VLT+FORS2        &  GRIS$\_$300V      &  13   \\   
20191011  & 58766.00  &  670.62  & 4750-9350 &  VLT+MUSE         &  WFM$^{\star}$     &  3    \\
\hline 
\end{tabular}
\begin{list}{}{}
\item \textbf{NOTES:}
\item $^{\ast}$ Rest-frame phase in days
from explosion.
\item $^{\star}$ MUSE was used in Wide Field Mode (WFM).
\item Tillinghast: 60in Tillinghast Telescope at F. L. Whipple Observatory (AZ, USA)
\item Ekar: Copernico 1.82m Telescope, INAF (Mt. Ekar, Asiago, Italy)
\item Pennar: Galileo 1.22m Telescope, DFA University of Padova (Asiago, Italy)
\item LT: 2.0m Liverpool Telescope (La Palma, Spain)
\item NTT: New Technology Telescope 3.6m, ESO (La Silla, Chile)
\item Keck: W. M. Keck Observatory, (Maunakea, Island of Hawai’i)
\item VLT: Very Large Telescope, ESO (Paranal, Chile)
\end{list}
\end{table*}

\renewcommand{\thetable}{A\arabic{table}}
\setcounter{table}{8}
\begin{table*}
\footnotesize
\centering
\caption{FWHM velocity and velocity offset for the H$\alpha$, \ion{O}{I}, \ion{[Fe}{II]} and \ion{[Ca}{II]} lines.}
\label{measurements}
\begin{tabular}[t]{c|cccc|cccc}
\hline
Phase$^{\ast}$ & \multicolumn{4}{c|}{FWHM velocity} & \multicolumn{4}{c}{Velocity offset} \\
\hline
&  H$\alpha$ 	&  \ion{O}{I}	& \ion{[Fe}{II]} & \ion{[Ca}{II]} & H$\alpha$ & \ion{O}{I} & \ion{[Fe}{II]} & \ion{[Ca}{II]} \\   
(days) & (km s$^{-1}$)& (km s$^{-1}$) & (km s$^{-1}$)  & (km s$^{-1}$)  & (km s$^{-1}$) & (km s$^{-1}$) & (km s$^{-1}$)  & (km s$^{-1}$) \\ 
\hline
110	& $6490\pm250$	& $4840\pm247$	& \nodata        & $3370\pm280$   &  $-1860$  &   $-1290$  &   \nodata	    & $-905$       \\
136	& $6110\pm290$	& $4570\pm224$	& $1815\pm240$   & $3470\pm265$   &  $-1860$  &   $-1385$  &   $-1925$      & $-1070$      \\
158	& $5970\pm270$	& $4965\pm271$	& $1910\pm285$   & $3370\pm240$   &  $-1630$  &   $-1195$  &   $-1925$      & $-1070$      \\
242	& $6770\pm380$	& $5520\pm224$	& $1955\pm495$   & $3230\pm240$   &  $-760$   &   $-620$   &   $-1590$      & $-990$       \\
249	& $6090\pm270$	& $4550\pm271$	& $1705\pm240$   & $3095\pm220$   &  $-715$   &   $-665$   &   $-1590$      & $-1155$      \\
255	& $7150\pm290$	& $5595\pm224$	& $1625\pm220$   & $3345\pm200$   &  $-625$   &   $-620$   &   $-1675$      & $-950$       \\
278	& $7525\pm315$	& $5895\pm484$	& $1765\pm370$   & $3295\pm220$   &  $-490$   &   $-335$   &   $-1590$      & $-950$       \\
285	& $7245\pm335$	& $5250\pm224$	& $1670\pm345$   & $3305\pm205$   &  $-535$   &   $-525$   &   $-1715$      & $-950$       \\
331	& $8235\pm405$	& $5355\pm224$	& $1675\pm325$   & $3345\pm225$   &  $-260$   &   $-380$   &   $-1590$      & $-990$        \\
337	& $7670\pm725$	& $5310\pm224$	& $1790\pm390$   & $3480\pm220$   &  $-260$   &   $-380$   &   $-1590$      & $-990$       \\
506	& $10035\pm475$	& \nodata	& \nodata        & $3410\pm200$   &  $-580$   &   $-240$   &   \nodata	    & $-1280$      \\
520	& $10460\pm430$	& $4695\pm508$	& \nodata        & $3715\pm200$   &  $-400$   &   $-145$   &   \nodata	    & $-990$       \\
573	& $10395\pm360$	& $5290\pm934$	& \nodata        & \nodata        &  $-535$   &   $-570$   &   \nodata	    & \nodata      \\
671	& $10510\pm290$	& $5895\pm1029$	& \nodata        & $4205\pm345$   &  $-400$   &   $-620$   &   \nodata	    & $-825$       \\
\hline 
\end{tabular}
\begin{list}{}{}
\item \textbf{Notes:} 
\item $^{\ast}$ Rest-frame phase in days from explosion. 
\end{list}
\end{table*}

\section{Figures: Gaussian fits}
\label{ap2}
\clearpage

\begin{figure*}
\centering
\includegraphics[width=16cm]{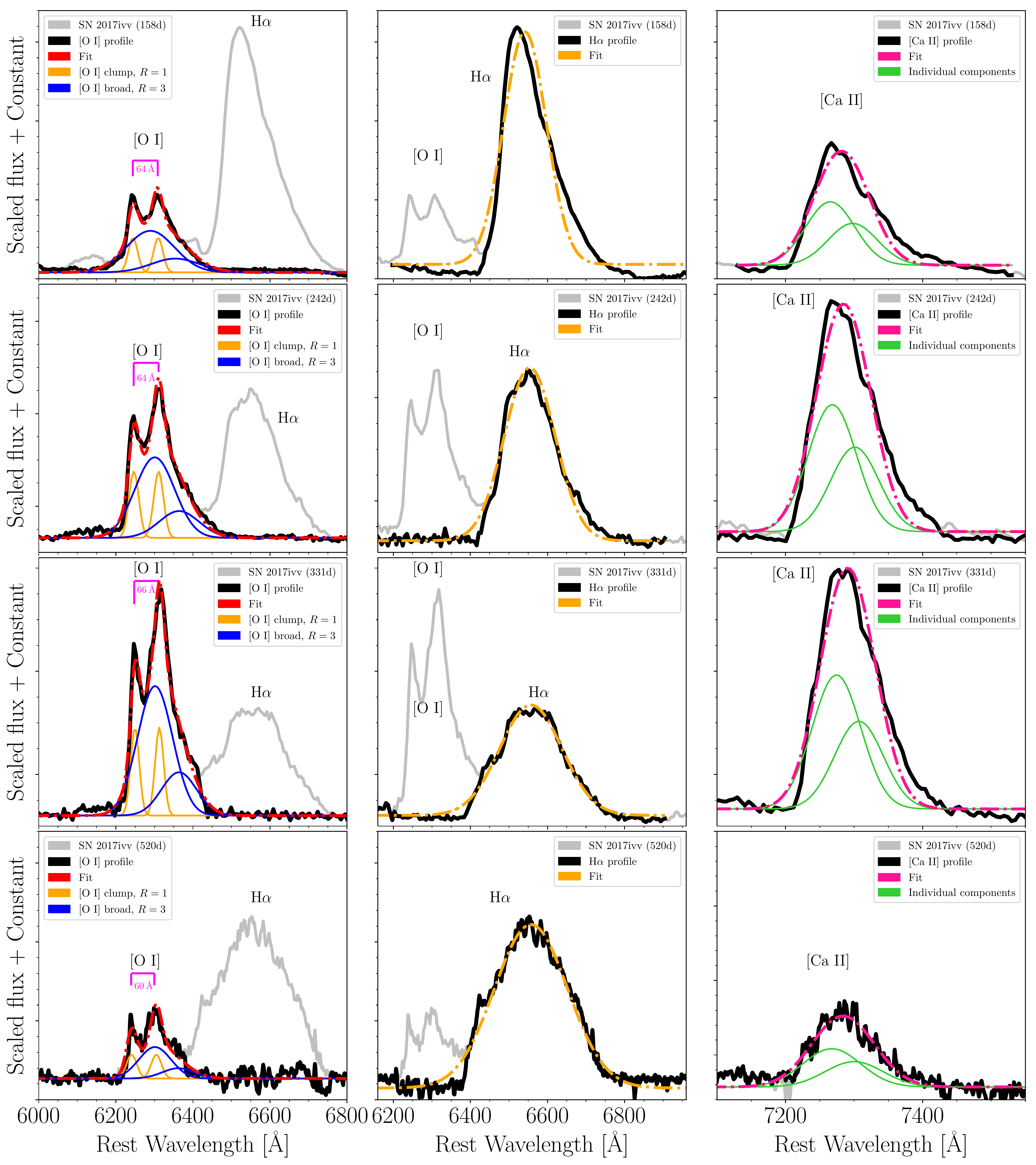}
\caption{Example of single Gaussian fits for \ion{[O}{I]} $\lambda\lambda6300$, 6364 (left panels), H$\alpha$ (central panels), and \ion{[Ca}{II]} $\lambda\lambda7291$, 7324 (right panels) at 158, 242, 331 and 520 days from explosion. On the top of the \ion{[O}{I]} $\lambda\lambda6300$, 6364 emission profile, the separation of the two peaks is marked in magenta.}
\label{gfits}
\end{figure*}
\clearpage

\parbox{\textwidth}{
$^{1}$ Department of Physics and Astronomy, University of Southampton, Southampton, SO17 1BJ, UK\\
$^{2}$ INAF - Osservatorio Astronomico di Padova, Vicolo dell'Osservatorio 5, I-35122 Padova, Italy\\
$^{3}$ The Oskar Klein Centre, Department of Astronomy, Stockholm University, AlbaNova, 10691 Stockholm, Sweden.\\
$^{4}$ Max-Planck-Institute for Astrophysics, 
Karl-Schwarzschild Strasse 1, 85748  Garching, Germany\\
$^{5}$ Departamento de F\'isica Te\'orica y del Cosmos, Universidad de Granada, E-18071 Granada, Spain\\
$^{6}$ European Southern Observatory, Alonso de C\'ordova 3107, Casilla 19, Santiago, Chile\\
$^{7}$ Max-Planck-Institut f{\"u}r Extraterrestrische Physik, Giessenbachstra\ss e 1, 85748, Garching, Germany\\
$^{8}$ Tuorla Observatory, Department of Physics and Astronomy, University of Turku, FI-20014 Turku, Finland\\
$^{9}$ Finnish Centre for Astronomy with ESO (FINCA), FI-20014 University of Turku, Finland.\\
$^{10}$ CENTRA, Instituto Superior T\'ecnico, Universidade de Lisboa, Av. Rovisco Pais 1, 1049-001 Lisboa, Portugal\\
$^{11}$ School of Physics \& Astronomy, Cardiff University, Queens Buildings, The Parade, Cardiff, CF243AA, UK\\
$^{12}$ School of Physics, O'Brien Centre for Science North, University College Dublin, Dublin, Ireland.\\
$^{13}$ School of Physics, Trinity College Dublin, The University of Dublin, Dublin 2, Ireland.\\
$^{14}$ Astrophysics Research Centre, School of Mathematics and Physics, Queens University Belfast, Belfast BT7 1NN, UK\\ 
$^{15}$ School of Physics and Astronomy, Tel Aviv University, Tel Aviv 69978, Israel\\
$^{16}$ CIFAR Azrieli Global Scholars program, CIFAR, Toronto, Canada\\
$^{17}$ INAF Osservatorio Astronomico di Padova, Vicolo dell'Osservatorio 5, 35122 Padova, Italy\\
$^{18}$ Astrophysics Research Institute, Liverpool John Moores University, 146 Brownlow Hill, Liverpool L3 5RF, UK\\
$^{19}$ Department of Astronomy, The Ohio State University, 140 W. 18th Avenue, Columbus, OH 43210, USA\\
$^{20}$ Center for Cosmology and AstroParticle Physics (CCAPP), The Ohio State University, 191 W. Woodruff Avenue, Columbus, OH 43210, USA.
$^{21}$ University of California, Davis\\
$^{22}$ Las Cumbres Observatory, Goleta, California 93117, USA\\
$^{23}$ Department of Physics, University of California, Santa Barbara, California 93106, USA\\
$^{24}$ Kavli Institute for Astronomy and Astrophysics, Peking University, Yi He Yuan Road 5, Hai Dian District, Beijing 100871, China \\
$^{25}$ Department of Astronomy, School of Physics, Peking University, Yi He Yuan Road 5, Hai Dian District, Beijing 100871, China \\
$^{26}$ The Oskar Klein Centre, Department of Astronomy, Stockholm University, AlbaNova, SE-10691 Stockholm, Sweden\\
$^{27}$ International Center for Relativistic Astrophysics, Piazzale della Repubblica 2, I-65122 Pescara, Italy\\
$^{28}$ Capodimonte Astronomical Observatory, INAF-Napoli, Salita Moiariello 16, 80131-Napoli, Italy \\
$^{29}$ Department of Particle Physics and Astrophysics, Weizmann Institute of Science, Rehovot 76100, Israel\\
$^{30}$ Astronomical Observatory, University of Warsaw, Al. Ujazdowskie 4, 00-478 Warszawa, Poland\\
$^{31}$ The Observatories of the Carnegie Institution for Science, 813 Santa Barbara St., Pasadena, CA 91101, USA\\
$^{32}$ Center for Astrophysics \textbar{} Harvard \& Smithsonian, 60 Garden Street, Cambridge, MA 02138-1516, USA\\
$^{33}$ Birmingham Institute for Gravitational Wave Astronomy and School of Physics and Astronomy, University of Birmingham, Birmingham B15 2TT, UK \\
$^{34}$ Institute for Astronomy, University of Edinburgh, Royal Observatory, Blackford Hill, EH9 3HJ, UK \\
$^{35}$ Departamento de Ciencias Fisicas, Universidad Andres Bello, Avda. Republica 252, Santiago, Chile\\
$^{36}$ Millennium Institute of Astrophysics (MAS), Nuncio Monse\~nor Sotero Sanz 100, Providencia, Santiago, Chile\\
$^{37}$ N\'ucleo de Astronom\'ia de la Facultad de Ingenier\'ia y Ciencias, Universidad Diego Portales, Av. Ej\'ercito 441 Santiago, Chile \\
$^{38}$ Institute for Astronomy, University of Hawai'i, 2680 Woodlawn Drive, Honolulu, HI 96822, USA\\
}

\bsp	
\label{lastpage}
\end{document}